\documentclass{article}
\usepackage{amsmath, amssymb}
\usepackage{graphicx}
\usepackage{graphics, epsfig,colordvi,afterpage}
\def\mm{{\mbox{max}}}
\def\be{\begin{equation}}
\def\ee{\end{equation}}
\def\la{\langle}
\def\ra{\rangle}
\def\peq{\stackrel{P}{=}}

\begin{document}
\bibliographystyle{plain}
\baselineskip = 24 pt

    \title{Primary visual cortex as a saliency map: parameter-free prediction of behavior from V1 physiology }
    \author{Li Zhaoping$^{1}$, Li Zhe$^2$}
    \date{$^1$University College London, UK, z.li@ucl.ac.uk; $^2$Tsinghua University, China.}
    \maketitle

\section*{Abstract} 

It has been hypothesized that neural activities in the primary visual  cortex (V1) represent a 
saliency map of the visual field to exogenously guide attention. This hypothesis has so 
far provided only qualitative predictions and their confirmations. 
We report this hypothesis' first quantitative prediction, derived without free parameters, 
and its confirmation by human behavioral data.
The hypothesis provides a direct link between V1 neural responses to a visual 
location and the saliency of that location to guide attention exogenously.
In a visual input containing many bars, one of them saliently different from all the other 
bars which are identical to each other,  saliency at the singleton's location can be 
measured by the shortness of the reaction time in a visual search task to find the singleton.
The hypothesis predicts quantitatively the whole distribution of the reaction times to find
a singleton unique in color, orientation, and motion direction from 
the reaction times to find other types of singletons.
The predicted distribution matches the experimentally observed distribution in
all six human observers. 
A requirement for this successful prediction is a data-motivated assumption 
that V1 lacks neurons tuned simultaneously to color, orientation, and motion direction 
of visual inputs.
Since evidence suggests that extrastriate cortices do have such neurons, 
we discuss the possibility that the extrastriate cortices play no role
in guiding exogenous attention so that they can be devoted to 
other functional roles like visual decoding or endogenous attention.

\vskip 0.3 in



\section*{Introduction}

Spatial visual selection, often called spatial attentional selection,
enables vision to select a visual location for detailed processing using limited cognitive
resources\cite{IttiKoch2001}. 
Metaphorically, the selected location is said to be in the attentional spotlight, which typically
coincides with the spatial zone centered on gaze position.
Hence, a visual input outside the spotlight, e.g., a letter in a word on this page
more than 10 letters from the current fixation location, is difficult to recognize.
Therefore, if one is to find a particular word on this page, the reaction time (RT) to find this 
word will depend on how long it takes the spotlight to arrive at the word location.
The spotlight can be guided by goal-dependent (or top-down, endogenous) mechanisms,
such as when we direct our gaze to the right words while reading, or by goal-independent (or bottom-up, exogenous) 
mechanisms
such as when we are distracted from reading by a sudden appearance of something in visual periphery.
In this paper, an input is said to be salient when it strongly attracts attention by bottom-up mechanisms,
and the degree of this attraction is defined as saliency.
For example, an orientation singleton such as a vertical bar in a background of horizontal bars 
is  salient, so is a color singleton such as a red dot among many green ones; and the location of such a 
singleton has a high saliency value.
Therefore, saliency of a visual location can often be measured by the shortness of the reaction time in 
a visual search to find a target at this location\cite{TreismanGelade80}, provided that 
saliency, rather than top-down attention, is the dictating factor to guide the attentional spotlight. 
It can also be measured in attentional (exogenous) cueing effect in terms of the degree in which 
a salient location speeds up and/or improves visual discrimination of a probe presented immediately 
after the brief appearance of the salient cue\cite{MullerRabbitt1989, NakayamaMackeben1989}.

Traditional views presume that higher brain areas such as those in the parietal and frontal parts of the brain
are responsible for saliency, i.e., to guide attention exogenously\cite{TreismanGelade80, DuncanHumphreys89, WolfeEtAl1989, IttiKoch2001}.
This belief was partly inspired by the observation that saliency is a general property that could 
arise from visual inputs with any kind of feature values (e.g., vertical or red) in any feature dimension (e.g., color, orientation, and motion) 
whereas each neuron in lower visual areas like the primary visual cortex is (more likely) tuned to specific feature values (e.g., a vertical orientation) 
rather than general visual features.
However, it was proposed a decade ago\cite{LiPNAS1999,  LiTICS2002} that the primary visual cortex (V1) computes
a saliency map, such that the saliency value of a location is represented by the 
highest response among V1 neurons to this location relative to the highest responses to the other 
visual locations, regardless of the preferred features of neurons giving such responses.
Although this V1 saliency hypothesis is a significant departure from traditional psychological
theories, it has received substantial experimental support\cite{ZhaopingSnowden2006, KoeneZhaoping2007, 
ZhaopingMay2007, JinglingZhaoping2008, Zhaoping2008OcularSingleton, ZhangZhaopingZhouFang2012, ZhaopingGaze2012}, detailed
in \cite{ZhaopingBook2014}.
In particular, behavioral data confirmed a surprising prediction from this hypothesis 
that an eye-of-origin singleton (e.g., an item uniquely shown to the left eye among other items shown to the right eye) 
that is hardly distinctive from other visual inputs can attract attention and gaze
qualitatively just like a salient and highly distinctive orientation singleton does.  In fact,
observations\cite{Zhaoping2008OcularSingleton, ZhaopingGaze2012} show  that an eye-of-origin singleton
can be even more salient than a very salient orientation singleton.
This finding provides a hallmark of the saliency map in V1
because the eye-of-origin feature is not explicitly represented in any visual cortical area except V1.
(Cortical neurons, except many in V1, are not tuned to eye-of-origin 
feature\cite{HubelWieselFerrierLecture1977, BurkhalterVanEssen1986}, making this feature non-distinctive 
to perception.)
Functional magnetic resonance imaging and event related potential measurements also
confirmed that, when top-down confounds are avoided or minimimzed, a salient location evokes
brain activations in V1 but not in the parietal and frontal regions\cite{ZhangZhaopingZhouFang2012}.

\begin{figure}[hhhhhhthh!!]
\begin{center}
\includegraphics[width=130mm]{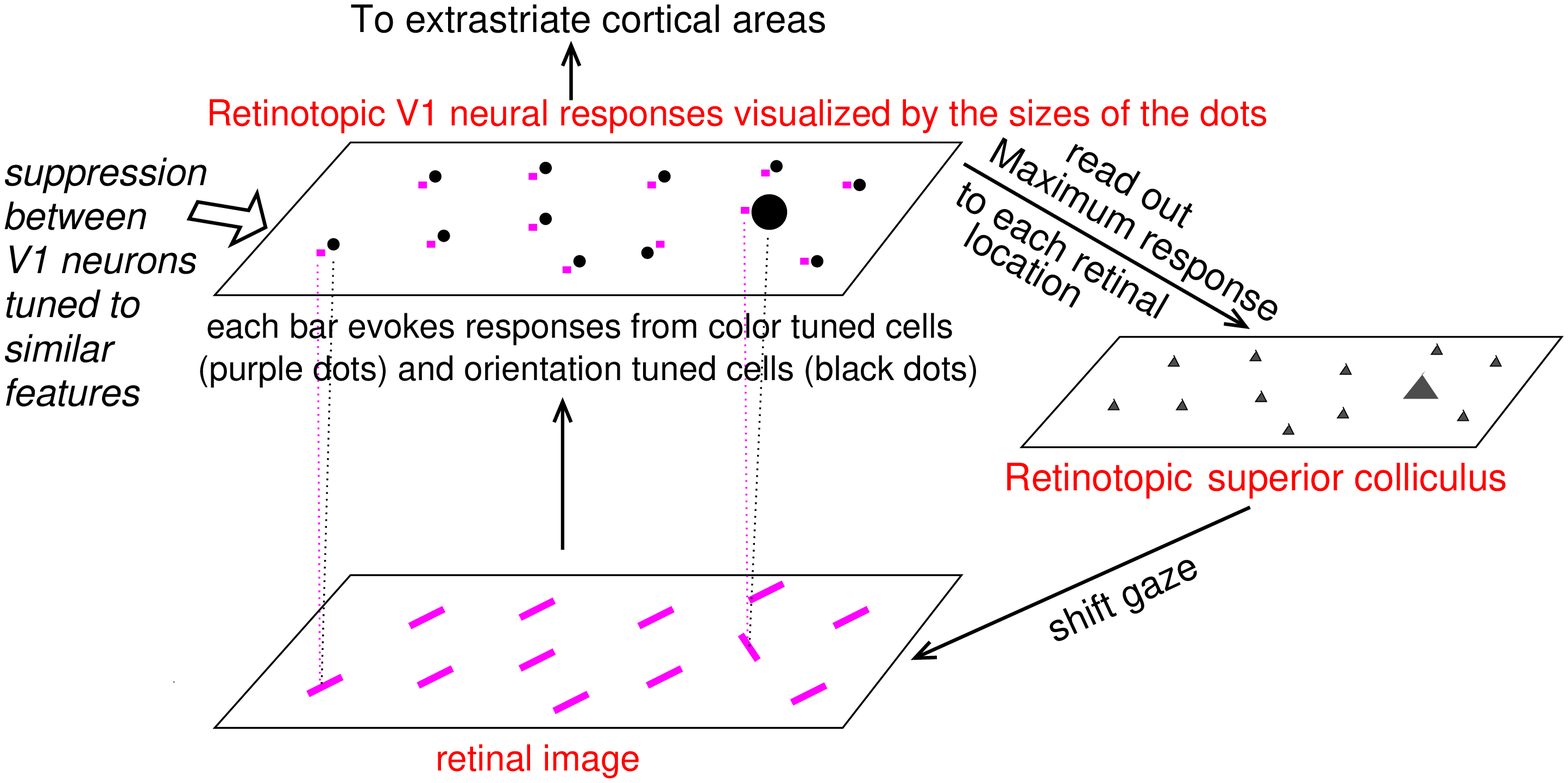}
\end{center}
\caption{ \label{fig:IntroA}
V1 saliency hypothesis states that the bottom-up saliency of a location is represented by the maximum V1 response to this location.
In this schematic for illustration, V1 is simplified to contain only two kinds of neurons, one tuned to color (their responses are
visualized by the purple dots) and the other tuned
to orientation (black dots).  Each input bar evokes responses in a cell tuned to its color and another
cell tuned to its orientation (indicated for two input bars by linking each bar to its two evoked responses by dotted lines),
and the receptive fields of these two cells cover the same retinal location even though (for better visualization) 
the dots representing these cells are not exactly overlapping in the cortical map.
Iso-feature suppression makes nearby V1 neurons tuned to similar features (e.g., similar color or similar orientation)
suppress each other. The orientation singleton in this image evokes the highest V1 response because the orientation tuned neuron responding
to it escapes iso-orientation suppression. The color tuned neuron tuned and responding to the singleton's color is under 
iso-color suppression. The saliency map is likely read out by the superior colliculus to execute gaze shifts to salient locations.
}
\end{figure}

So far, the existing tests of the V1 saliency hypothesis have been qualitative. 
Here, we report its first quantitative prediction that is derived without free parameters. 
This prediction is of the distribution of the reaction times in a visual search for a
singleton bar defined by its uniqueness in color, orientation, and motion direction
among uniformly featured background bars.  This prediction can be made because the hypothesis  
directly links the response properties of V1 neurons with the reaction times of visual searches.
Specifically, according to the hypothesis, the saliency of a visual location is represented by the
maximum of the responses of V1 neurons to this location, regardless of the input feature
selectivity of the neurons concerned\cite{LiPNAS1999,LiTICS2002}.
For example, a visual input in Fig. \ref{fig:IntroA} contains many colored bars, each
activates some V1 neurons tuned to its color and/or orientation. The highest response to each bar
signals the saliency of its location according to the V1 hypothesis, regardless of whether the V1 neuron
giving this (highest) response is tuned to the color or orientation (or both color and orientation) of the bar. 
These highest V1 responses for various visual locations thus represent a saliency map of  the scene.
This saliency map may potentially be read out by the superior colliculus, which receives mono-synaptic input 
from V1 and controls eye movement to execute the attentional selection\cite{Schiller1998}.
If an observer searches for a uniquely oriented bar in the retinal image in Fig. \ref{fig:IntroA}, 
the reaction time to find this bar, associated with the saliency of the target location, should thus be associated with 
the highest V1 response to the target. 
In particular, a shorter reaction time should result from a larger value of the highest
response to the search target (when the highest responses to various 
non-target locations are fixed).  
As will be explained in detail, a feature singleton, e.g., an orientation singleton, 
tends to be the most salient in a scene because it tends to evoke the highest V1 response to 
the scene due to iso-feature suppression\cite{LiPNAS1999}, the mutual suppression between nearby 
neurons preferring the same or similar features\cite{AllmanEtAl1985}: iso-feature suppression makes 
neurons responding to a non-singleton item suppressed by other neurons
responding to neighboring non-singletons sharing the same or similar features.
As will be shown, the direct link between the reaction times and V1 responses assumed by the V1 saliency hypothesis,
together with V1's neural response properties 
(in particular iso-feature suppression and feature selectivities by the neurons), 
enables a quantitative prediction on the distribution of the reaction times without any free parameters.
Furthermore, we will show that this prediction matches behavioral reaction time data quantitatively.

In addition, this paper explores the implications of the confirmation of this quantitative prediction 
by experimental data.  
We will show that the prediction arises when the cortical area(s) responsible for computing saliency 
satisfies two requirements, one functional and one physiological.
The functional requirement is, as stated by the V1 saliency hypothesis, that the saliency 
of a location is signalled by the highest response to that location among the responses 
from the cortical neurons.
The physiological requirement is that the saliency computing cortical area(s) 
should have the following properties found in V1:  a neuron's response should be tuned to color, 
orientation, or motion direction, or tuned simultaneously to any two of these three feature dimensions;  
however, there should  be few neurons tuned simultaneously to all the three feature 
dimensions\cite{HubelWieselFerrierLecture1977, LivingstoneHubel1984, HorwitzAlbright2005} with the ubiquitously 
associated iso-feature suppression.
Hence, the confirmation of the prediction enables us to
identify possible candidate brain areas for saliency computation.
In principle, if an extrastriate area also satisfies the physiological requirement, 
it might also play a role in saliency together with V1. 
We will discuss experimental evidence on whether the extrastriate cortical areas satisfy 
this physiological requirement and thus whether they can be excluded from playing 
a role in saliency. Parts of this work have been presented in abstract 
form elsewhere\cite{ZhaopingZheVSS2012, ZhaopingECVPSpeech2013}.

\section*{Results}

In this section, starting from an overview of the background of V1 mechanisms and the V1 saliency hypothesis, 
we show a direct link between the reaction time to find a visual feature singleton in a homogeneous background
(like that in Fig. \ref{fig:IntroA}) and the highest V1 response to this singleton.  
From this link, we derive the quantitative prediction of the hypothesis 
and present its experimental test using behavioral data.
In this process,  we also present some related but spurious theoretical predictions that should be violated
unless certain conditions on the V1 neural mechanisms hold. 
These spurious predictions and their tests (falsification) by behavioral reaction time data
not only help to provide further insights in the underlying neural mechanisms but also help to illustrate 
and verify our methods.

\subsection*{Iso-feature suppression between neurons as the mechanism for high saliencies of feature singletons}

In the retinal image of Fig. \ref{fig:IntroA}, the location of an orientation singleton, a left-tilted bar in a background of 
right tilted bars, is most salient. This is because a V1 neuron tuned to its orientation, with its receptive field 
covering the bar, responds more vigorously than any neuron responding to the background bars. 
Note that, throughout the paper, `a neuron responding to a bar' means the most responsive neuron among a 
local population of neurons with similar input selectivities responding to this bar regardless of the number
of neurons in this local population. The higher response to the orientation singleton is due to iso-orientation 
suppression between nearby neurons tuned to same or similar orientations\cite{AllmanEtAl1985,KnierimVanEssen1992, LinLi1994}.
Hence, neurons responding to neighboring background bars suppress each other because they are tuned to the
same or similar orientation, whereas the neuron responding to the orientation singleton escapes such suppression
because it is tuned to a very different orientation. 

In addition to the orientation feature, V1 neurons are also tuned to other input feature dimensions including
color,  motion direction, and eye of origin\cite{HubelWiesel1968, LivingstoneHubel1984}. 
Hence, each colored bar in the retinal image of Fig. \ref{fig:IntroA} evokes not only a response in a cell tuned
to its orientation but also another response in another cell tuned to its color (omitting other input features for simplicity), 
this is indicated by the dotted lines linking the two example input bars and their 
respective evoked V1 responses.  In general, there are many V1 neurons whose receptive fields cover the location of 
each visual input item (including neurons whose preferred orientation or color does not match the visual input feature), 
and only the highest response from these neurons represents the saliency of this location according to the 
V1 saliency hypothesis (note that this highest response is unlikely to be from a neuron whose preferred feature
is not in the input item). In the example of Fig. \ref{fig:IntroA},
responses from the color tuned neurons to all bars suffer from iso-color suppression\cite{WachtlerEtAl2003}, 
which is analogous to iso-orientation suppression,  since all bars have the same color.
Focusing on V1 neurons tuned to color only and neurons tuned to orientation only for simplicity, 
the highest response evoked by the orientation singleton is in the orientation-tuned rather than
the color-tuned cell, and this response alone (relative to the responses to the background bars) determines the saliency of 
the orientation singleton.
Later in the paper, the notion that many V1 neurons respond to a single
input location or item will be generalized to include neurons tuned to motion direction and neurons 
jointly tuned to multiple feature dimensions. Determining the highest V1 response to each input location 
will involve determining which of the many neurons whose receptive fields cover this location 
has the highest response. 

Analogous to iso-orientation suppression and iso-color suppression, iso-motion-direction and iso-ocular-origin suppressions
are also present in V1\cite{AllmanEtAl1985, KnierimVanEssen1992, LinLi1994, DeAngelisEtAl1994, 
WachtlerEtAl2003, JonesEtAl2001}, and we call them iso-feature suppression in general\cite{LiPNAS1999}.
Accordingly, an input singleton in any of these feature dimensions should be salient (see Fig. \ref{fig:IntroB}B
for a color singleton), since the neuron responding to the unique feature of the singleton escapes the 
iso-feature suppression of the neurons responding to the uniformly featured background items.
This is consistent with known behavioral saliency and has led to the successful prediction of the salient 
singleton in eye-of-origin\cite{Zhaoping2008OcularSingleton}.
Iso-feature suppression is believed to be mediated by intra-cortical 
neural connections\cite{RocklandLund1983,GilbertWiesel1983} linking neurons whose receptive fields
are spatially nearby but not necessarily overlapping.

\subsection*{The feature-blind nature of saliency representation in V1}

\begin{figure}[hhhhhhthh!!]
\begin{center}
\includegraphics[width=130mm] {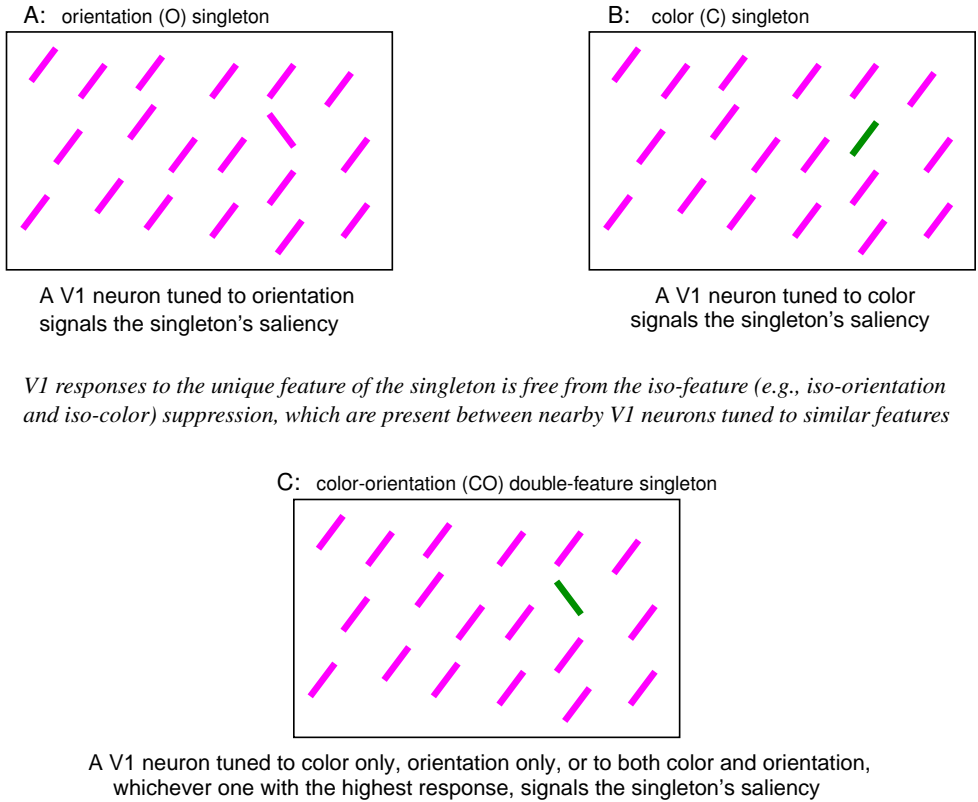}
\end{center}
\caption{ \label{fig:IntroB}
Schematics of visual stimuli for singleton searches.
Due to iso-feature suppression, the most active response to each image is from a neuron responding to the singleton bar.
This most activated neuron is  tuned to orientation in image A, tuned to color in image B, and to color, orientation, or both features of the singleton 
for image C.
The highest V1 response to the singleton signals the saliency of its location.
}
\end{figure}

According to the V1 saliency 
hypothesis, it is only V1 response vigor that matters for saliency, and not the visual feature value  concerned. 
Let us compare Fig. \ref{fig:IntroB}A and Fig. \ref{fig:IntroB}B: one has an orientation (O) singleton and
one has a color (C) singleton, and they share the same background bars.
In each image, the singleton should activate some neurons which are orientation tuned and some other
neurons which are color tuned (for the moment, we omit for simplicity neurons tuned simultaneously to color and orientation).  
In Fig. \ref{fig:IntroB}A, 
the most activated neuron by the singleton is orientation tuned due to iso-orientation suppression; 
color tuned neurons responding to any bar, singleton or otherwise, suffer iso-color suppression.
In Fig. \ref{fig:IntroB}B, the most activated neuron is color tuned instead; and the orientation 
tuned neurons responding to any bar suffer iso-orientation suppression.
However, if the highest responses evoked by the two singletons are identical, 
then the two singletons are equally salient (assuming that the population responses to the background bars in the
two images are identical), even though different singletons evoke this highest response in neurons tuned to different feature dimensions.
Conversely, if the respective highest responses evoked by the two singletons are different from each other, 
the singleton evoking the higher response is more salient, regardless of neurons giving the highest responses.
The feature-blind nature of this saliency representation in V1 enables
the brain to have a bottom-up saliency map in V1, despite the feature tuning of V1 neurons, without resorting to
higher cortical areas such as the frontal eye field or lateral-intraparietal cortex\cite{GottliebEtAl1998,IttiKoch2001}.

\subsection*{Linking V1 responses with reaction times}

When the effect of top-down attentional guidance is negligible in a visual search task, 
a higher saliency at the target location should lead to a shorter reaction time to find the target, 
by the definition of saliency. In stimuli like those in Fig. \ref{fig:IntroB},  the feature singletons are
so salient that we can assume that its saliency dictates the immediate attention shifts upon the appearance of the
stimuli.  It is typical that  the first saccade after the appearance of such search images is
directed to the feature singleton.  However, the latency of this attentional shift is shorter for a more salient 
singleton. Assuming a fixed additional latency from the shift of attention to the singleton to an observer's response
to report the singleton, then the reaction time for the visual search task is determined by the singleton's saliency.

Let a visual image (or scene)  have visual input items at $n$ locations $i = 1, 2, ..., n$,  
and let $r_i$ be the highest V1 response (among multiple responses from multiple V1 neurons) 
evoked by location $i$. Then the saliency of location $i$ is 
determined by the value of $r_i$ relative to other values $r_j$ for $j\ne i$.
This is because, according to the V1 saliency hypothesis,  saliency read-out process is like an auction for 
attention, with $r_i$ the bidding price for attention by location $i$, such that the location giving the highest bid 
is the most likely to win attention\cite{ZhaopingNetworkReview2006}. Let us order $r_i$ such that
\begin{equation}
r_1 \ge r_2 \ge r_3 \ge ... \ge r_n, 
\end{equation}
then the first location is the most salient in the scene. Formally, we can use a function
$g(.)$ to describe
\begin{equation}
	\mbox{saliency at the most salient location} = g(r_1 | r_2, r_3, ..., r_n). \label{eq:GeneralG}
\end{equation}
In this paper, we are concerned only with visual scenes like those in  Fig. \ref{fig:IntroB}.
Each of such a scene is called a feature singleton scene in this paper. It has one feature singleton 
in a background of many items which are identical to each other, and the feature singleton is
far more salient than all other input items.  Then, $r_1$ is the highest response evoked by the singleton and is 
substantially and significantly larger than any $r_{i}$ for $i>1$.  For example,  the background bars 
may evoke $r_i$ (for $i>1$) that are no more than 10 spikes/second whereas the singleton evokes a $r_1$ that is no less 
than 20 spikes per second.
In such a feature singleton scene when $n$ is very large (e.g., 660 in the visual stimuli we will use later), 
we can reasonably expect that $g(r_1|r_2, r_3, ...)$,
the saliency of the singleton,  depends on $(r_2, r_3, ... )$ mainly through the statistical properties across the
$r_i$'s rather than the exact value of each $r_i$ for $i>1$.
For example, the statistical properties of $(r_2, r_3, ..., r_n)$ can be partly characterized by the 
average $\bar r$ and standard deviation $\sigma$ across $r_i$'s for $i>1$; then a singleton with a larger
$(r_1 -\bar r)/\sigma $, and perhaps also a larger $r_1/\bar r$, 
tends to be more salient\cite{LiPNAS1999}. 
More strictly, the function $g(.)$ in equation (\ref{eq:GeneralG}) may also depend on the locations $x_i$ of visual inputs
associated with $r_i$ for all $i$. However, this paper assumes that this dependence is negligible when
we restrict our visual scenes to the singleton scenes satisfying the following: (1) the eccentricity of the singleton from the
center of the visual field is fixed, (2) different $r_i$ and $r_j$ are sufficiently similar for all $i>1$ and $j>1$ ($j\ne i$) and 
the spatial distribution of the locations of the non-singleton items (whose $r_i$ are those with $i>1$) 
is approximately fixed. This paper is concerned only with scenes which are assumed to satisfy these conditions. 

If a set of visual scenes  are such that all scenes in this set are 
identical in terms of the number $n$ of visual input locations in each scene and the distribution of the 
response values $r_i$  for $i>1$, then we say that these scenes share 
an {\it invariant background response distribution}. 
For example, the three feature singleton scenes in Fig. \ref{fig:IntroB} can be approximately seen as to share 
an invariant background response distribution, even though the highest response $r_1$ to the singleton may be larger in  Fig. \ref{fig:IntroB}C than Fig. \ref{fig:IntroB}AB. 
This is because the responses to each background bar in each scene
are determined by the direct visual input (the bar itself) and the neighboring bars which exert contextual 
influence (mainly iso-feature suppression).
The higher responses to the feature singleton
should have a relatively small or negligible influence on the responses to the background bars due to a reduction or
elimination of iso-feature suppression between neurons responding to different features.
In any case, most contextual neighbors of each background bar 
are the background bars, not the feature singleton. 
The properties of the invariant background response distribution, or the statistical properties of 
the (highest) responses $(r_2, r_3, ..., r_n)$ to the background bars, are determined by such characteristics 
as the density, contrast, and the degree of regularities in the spatial placements of the background bars.

Therefore, given a fixed invariant background response distribution shared by a set of feature singleton scenes,  
we can assume that the saliency of the singleton can be approximately seen as depending only on the highest 
response $r_1$ to the
singleton.
Then, we can omit the explicit expression of $(r_2, r_3, ... )$ in equation (\ref{eq:GeneralG}) and write 
(still using the same notation $g(.)$ for convenience)
\begin{equation}
        \mbox{the saliency of the singleton location} = g(\mbox{$r_1$, the highest response to the singleton}).  \label{eq:SingletonSaliency}
\end{equation}
The $g(r)$ monotonically increases with $r$, and its exact dependence on $r$ is determined by the
properties of the invariant background response distribution.  
Since a larger saliency at the location of the singleton should give a 
shorter reaction time to find it (assuming again top-down factors are negligible), we can 
write this reaction time also directly as a function of the highest response to the singleton:
\begin{equation}
        \mbox{the reaction time to find a feature singleton} = f(\mbox{$r_1$, the highest response to the singleton}),  \label{eq:SingletonRT}
\end{equation}
in which $f(.)$ is a monotonically decreasing function of its argument. 
The exact form of $f(.)$ should depend on the invariant background response distribution and on the saliency 
read-out system. It can also depend on the observer 
(e.g., some observers can respond faster than others). We will see that these details about $f(.)$ do not matter in
our study.  Regardless of these details, among feature singleton scenes sharing an invariant background response distribution,   
two singletons evoking the same highest V1 response should require the same
reaction time to find them for a given observer, at least statistically, and the singleton evoking a larger V1 response (the highest response)  
should require a shorter reaction time. With this, the reaction time of a visual search for a feature singleton is directly linked to the V1 responses.

\subsection*{A race model}

Let us apply equation (\ref{eq:SingletonRT}) to the singleton scenes like those in 
Fig. \ref{fig:IntroB}, when these scenes share an invariant background response distribution.
For ease of argument, we start first by a simplified toy V1 which is assumed to have
only two kinds of neurons, one tuned to color only and one tuned to orientation only. 
This assumption is untrue in the real V1; we make it temporarily in this toy V1 to illustrate the method.
Furthermore, we assume that V1 responses are deterministic rather than stochastic given a visual input. 
(These simplifications will be removed later.)
Let $r^O$ or $r^C$, respectively, denote the response of the orientation tuned neuron  or the color tuned neuron
to the singleton in Fig. \ref{fig:IntroB}A or Fig. \ref{fig:IntroB}B, respectively.
Due to iso-feature suppression, 
$r^O$ and  $r^C$
are also the highest responses to the respective singletons.
Let ${RT}_O$ and ${RT}_C$ denote the reaction times to find the
orientation and color singletons, respectively. Then, 
\begin{equation}
{RT}_O = f(r^O)     \quad\mbox{and} \quad {RT}_C  = f(r^C) \label{eq:RTOandRTC}
\end{equation}
according to equation (\ref{eq:SingletonRT}).

Consider now the case that  the singleton bar is unique in both orientation and color, as
in Fig \ref{fig:IntroB}C.  This singleton is a double-feature singleton, while the 
singletons in Fig \ref{fig:IntroB}AB are single-feature singletons.
By iso-feature suppression, both the neuron tuned to the unique orientation and the
neuron tuned to the unique color will be more vigorously activated than neurons responding to 
the orientation and color of the background bars.  
Furthermore, we assume that the response of the orientation tuned 
neuron (and the contextual influences on it) should not be affected by the color of the input such that
the response $r^O$ to the singleton should be identical in Fig \ref{fig:IntroB}A and Fig \ref{fig:IntroB}C. 
Analogously, the response $r^C$ of the color tuned neuron to the singleton should be identical in Fig \ref{fig:IntroB}B and Fig \ref{fig:IntroB}C.
Hence, the maximum V1 response to the singleton in Fig \ref{fig:IntroB}C is max$\left (r^C, r^O \right )$ (where max(.) means to take
the maximum value among the arguments), and 
the reaction time ${RT}_{C\!O}$ to find the double-feature singleton is 
\begin{equation}
	{RT}_{C\!O} = f\left [\mbox{max}\left (r^C, r^O \right ) \right ] = \mbox{min}\left [f\left (r^C \right ), f\left (r^O \right ) \right ] = \mbox{min}({RT}_C, {RT}_O),  \label{eq:COraceDeterministic_V1}
\end{equation}
when we combine equations (\ref{eq:SingletonRT}) and (\ref{eq:RTOandRTC}) and the fact that $f(.)$ is a monotonically
decreasing function (min(.)  means to take the minimum value of the arguments). 

Hence, in the toy V1 which has only neurons tuned to orientation only and neurons tuned to color only but no neurons tuned to both, 
the V1 saliency hypothesis predicts that the double-feature singleton should be as salient 
as the more salient of the two single-feature singletons, such that the reaction time to find the CO singleton
is the shorter one of the reaction times for the single-feature singletons. 
If for example ${RT}_C = 400 $ millisecond (ms) and ${RT}_O = 500$ ms, then ${RT}_{C\!O} = 400$ ms. 
The equation 
\begin{equation}
{RT}_{C\!O} = {\rm min}({RT}_C, {RT}_O) \label{eq:COraceDeterministic}
\end{equation}
describes the deterministic version of a race model\cite{Raab1962} often used
to model a behavioral reaction time as the shorter reaction time of 
two or more underlying processes with their respective reaction times. 
For our example, it is as if the reaction time for the CO singleton is the winning
reaction time in a race between two racers with their respective reaction times.
We note that this race model equation does not depend on the detailed form of saliency read-out function $f(.)$ 
as long as $f(.)$ is a monotonically decreasing function.

As V1 responses are actually stochastic,  responses $r^C$ and $r^O$ to the feature singletons 
in Fig \ref{fig:IntroB} and the responses to the background items are all random samples from their respective distributions.
Despite this stochasticity, we assume the following two conditions hold.
First, the number $n$ of the background items is sufficiently large such that the statistical 
properties of the invariant background response distribution (e.g., the mean and standard deviation across the responses to 
the background items) are unchanged, or are not stochastic, despite the stochasticity of responses to individual background items. 
Second, the singletons are salient enough that the responses
$r^C$ and $r^O$ to the feature singletons are always the highest
responses to their respective scenes.
Consequently, equation (\ref{eq:RTOandRTC}) still holds,  and the stochasticity of
$r^C$ and $r^O$ simply means the corresponding 
stochasticity in ${RT}_C$ and ${RT}_O$, respectively.
For example, if $P_{r^O}(r^O)$ is the probability density of $r^O$, then the probability density of ${RT}_O$ is
\begin{equation}
	P_{{RT}_O} ({RT}_O) = \left [P_{r^O}(r^O) \left ({{df(r^O)}\over {dr^O}}\right )^{-1} \right ], \quad \mbox{at}\quad r^O = f^{-1}({RT}_O).
\end{equation}  
In any case, ${RT}_{C\!O} = f\left [{\rm max}\left (r^C, r^O \right ) \right ] = {\rm min}\left [f\left (r^C), f(r^O\right )\right ]$  still holds. 
If $r^C$ and $r^O$ fluctuate independently of each other 
for the responses to the double-feature singleton, then the stochastic version of the race model (in equation (\ref{eq:COraceDeterministic})) is
\begin{equation}
\mbox{Distribution of~} {RT}_{C\!O} = \mbox{Distribution of ~} \mbox{min} ({RT}_C, {RT}_O), 
 \label{eq:COracePrediction}
\end{equation}
in which ${RT}_C$ and ${RT}_O$ are independent random samples from their respective distributions.
For example, if the average of  ${RT}_C$ and ${RT}_O$ are 400 and 500 ms, respectively, the average of
${RT}_{C\!O}$ will be shorter than 400 ms by this stochastic race model, since each sample of ${RT}_{C\!O}$ 
is the race winner of the two samples ${RT}_C$ and ${RT}_O$.
This reflects statistical facilitation in this race model between the two single-feature 
singletons. 
For simplicity, we use 
\begin{equation}
{RT}_{C\!O}~\peq~\mbox{min} ({RT}_C, {RT}_O) \label{eq:COracePrediction_SH}
\end{equation}
as a shorthand for equation (\ref{eq:COracePrediction}), with the notation $x ~\peq~ y $ to 
mean that $x$ and $y$ have the same probability distribution. 

The race model, or race equality,  ${RT}_{C\!O} \peq \mbox{min} ({RT}_C, {RT}_O)$ is a prediction of the V1 saliency 
hypothesis if one were hypothetically to assume a toy V1 in which there is no V1 neuron which can respond more 
vigorously to the double-feature singleton than the orientation-only  tuned neuron and the color-only  tuned neuron. 
This assumption is wrong, even though it enables us to predict the distribution of 
${RT}_{C\!O}$ from those of ${RT}_C$ and ${RT}_O$.
Next we show that the predicted distribution of ${RT}_{C\!O}$ does not agree with the
behavorial data previously collected by Koene and Zhaoping\cite{KoeneZhaoping2007} (see Methods section).

\begin{figure}[hhhhhhthh!!]
\begin{center}
\includegraphics[width=90mm]{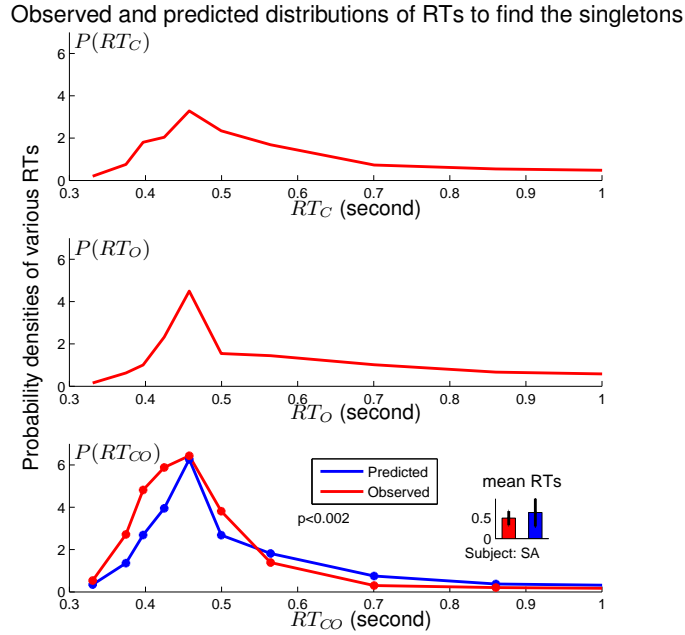}
\end{center}
\caption{ \label{fig:COSearch}
Behavioral refutation of a spurious prediction ${RT}_{C\!O}\peq {\rm min}({RT}_C, {RT}_O)$ 
based on the incorrect assumption that V1 lacks neurons tuned simultaneously to both orientation and color. 
The graphs show distributions (in discrete time bins) 
of ${RT}_O$, ${RT}_C$, 
and ${RT}_{C\!O}$ (and the average and the standard deviation of ${RT}_{C\!O}$) 
of a particular observer SA in searches of the singletons. 
There were respectively 296, 306, and 308 behavioral data samples 
for ${RT}_O$, ${RT}_C$, and ${RT}_{C\!O}$.  
Experimental data are shown in red, the prediction is in blue.
The predicted and actual distributions of ${RT}_{C\!O}$ are significantly different from each other, as indicated
by a $p<0.002$ in the bottom plot.
}
\end{figure}

\subsection*{The spurious race equality ${RT}_{C\!O} \peq {\rm min}({RT}_C, {RT}_O)$ is violated}

Figure \ref{fig:COSearch} compares the distribution of the behavioral data ${RT}_{C\!O}$ with the
distribution predicted from the behavorial data of ${RT}_C$ and ${RT}_O$ 
using the race equality ${RT}_{C\!O} \peq {\rm min}({RT}_C, {RT}_O)$. 
Statistical facilitation by the race model makes the predicted ${RT}_{C\!O}$ more densely 
populated in the shorter reaction time regions than the racers ${RT}_C$ and ${RT}_O$.  
Nevertheless, the behavioral ${RT}_{C\!O}$s are even shorter than the predicted ones.
The predicted and the actual ${RT}_{C\!O}$ distributions are 
significantly different from each other ($p<0.002$).  
(See the Methods section for the detailed procedures to predict the distribution of ${RT}_{C\!O}$ from those 
of ${RT}_C$ and ${RT}_O$ and to test the statistical significance of the difference between the 
predicted and observed distributions of ${RT}_{C\!O}$.)

\begin{figure}[tttt!]
\begin{center}
\includegraphics[width=140mm]{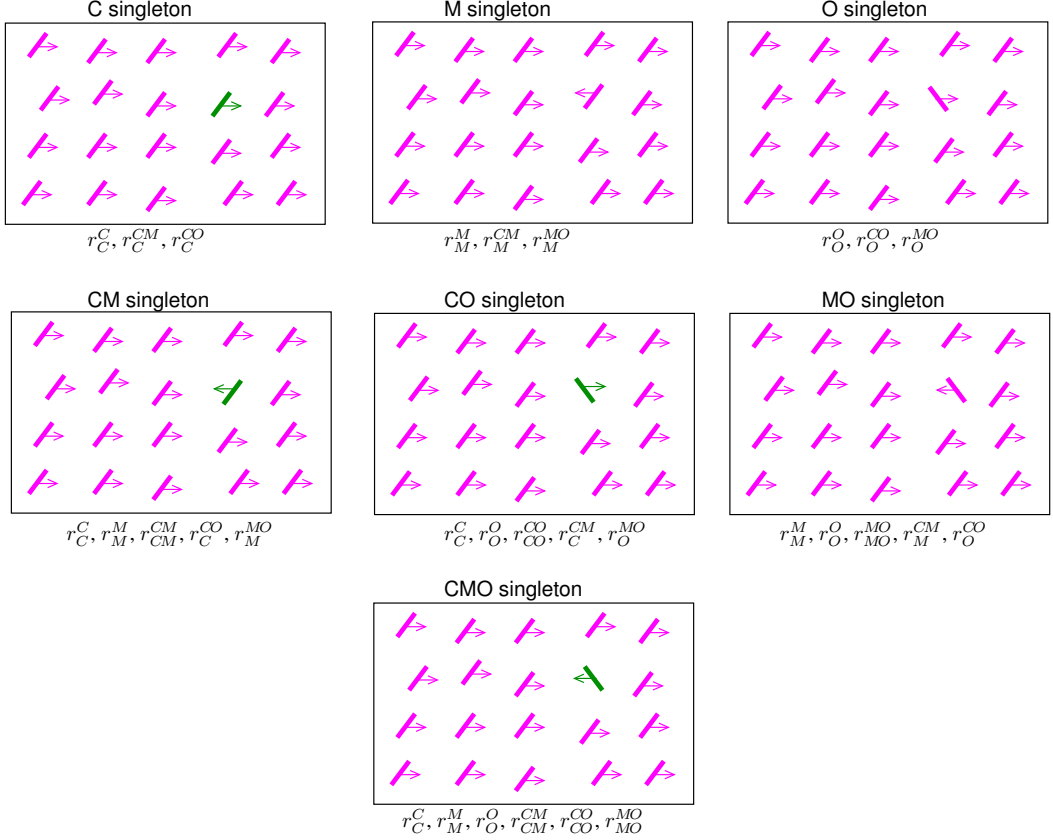}
\end{center}
\caption{\label{fig:SevenSingletons}
Schematics of the seven kinds of feature singletons. Each bar is colored green or purple (of the same luminance in the behavioral
experiment), tilted to the left or right from vertical by the same absolute tilt angle, moving to the left or right by the same motion speed.
The motion direction is schematically illustrated by an arrow pointing to the left or right.
Under each schematic, the non-trivial neural responses (e.g., these responses are expected to 
be substantially higher than the responses to the background bars) evoked by the singleton are listed.
}
\end{figure}

\begin{figure}[hhhhhhthh!!]
\begin{center}
\includegraphics[width = 145mm]{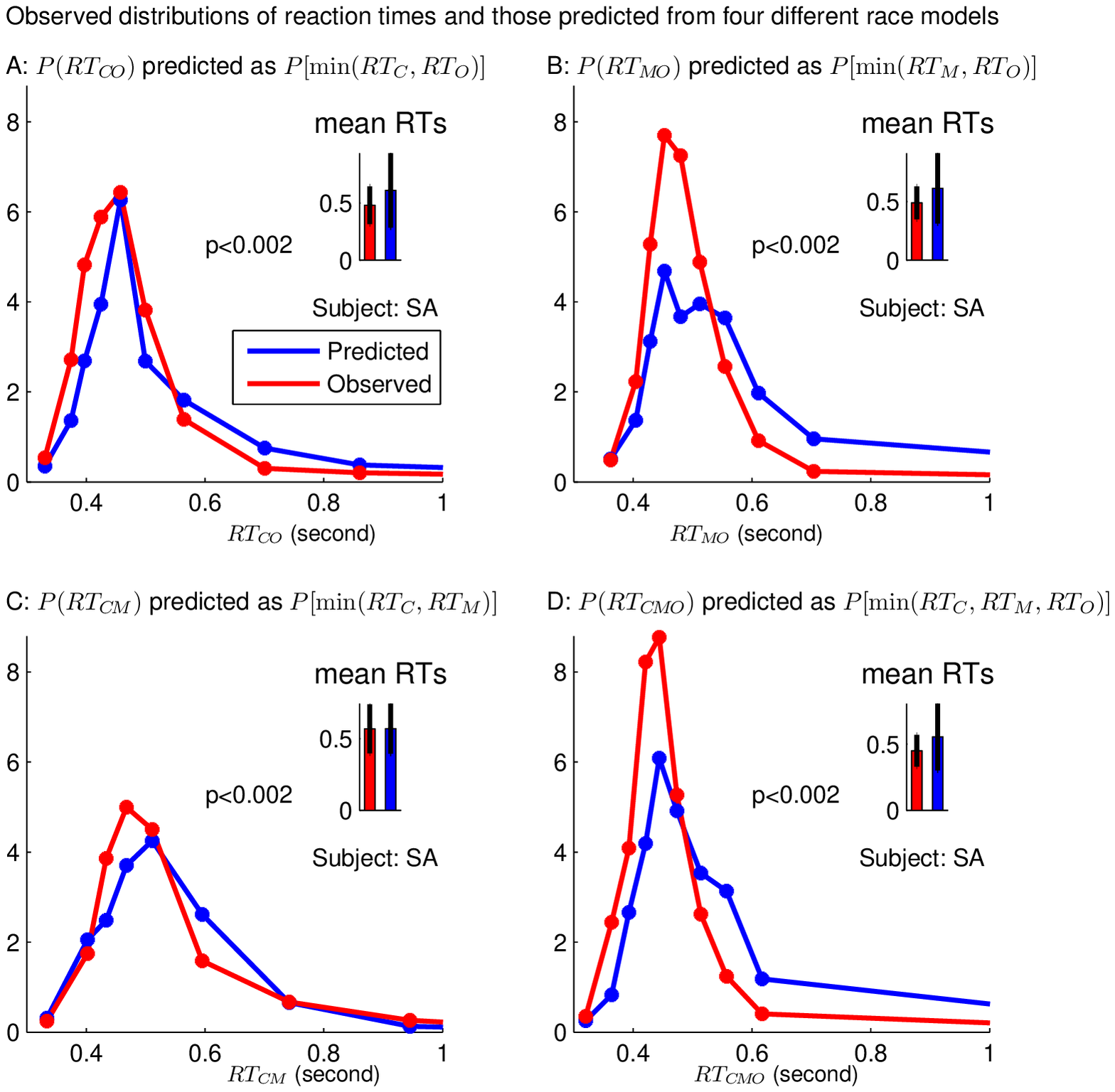}
\end{center}
\caption{ \label{fig:OneSubject_Race5to8}
The observed and predicted distributions of reaction times for a double- or triple-feature singleton, using four different race models (race equalities), 
${RT}_{C\!O} \peq  {\rm min}({RT}_C, {RT}_O)$ (in panel A), 
${RT}_{M\!O} \peq  {\rm min}({RT}_M, {RT}_O)$ (in panel B), 
${RT}_{C\!M} \peq  {\rm min}({RT}_C, {RT}_M)$ (in panel C), 
or ${RT}_{C\!M\!O} \peq  {\rm min}({RT}_C, {RT}_M, {RT}_O)$ (in panel D), 
in a race between the corresponding racers whose reaction times are those of the corresponding single-feature singletons.
The data is from the same subject SA already shown in Fig. \ref{fig:COSearch}, 
panel A shows the same information as that in the bottom panel of  Fig. \ref{fig:COSearch}.
In each panel, the distributions of the predicted and the observed reaction times, 
respectively, are significantly different from each other.
}
\end{figure}

When we include motion direction as an additional visual feature, a feature singleton in motion 
direction (M) is the analogy of a C or O singleton.  
Similarly, analogous to a CO singleton, a double-feature singleton CM or MO is unique in both color and motion 
direction, or in both motion direction and orientation, respectively. 
A triple-feature CMO singleton is unique all the three feature dimensions.
Fig. \ref{fig:SevenSingletons} shows the schematics of all the seven types of singletons. 
Let the reaction times to find singletons M, CM, MO, and CMO  be $RT_M$, $RT_{C\!M}$,  $RT_{M\!O}$, and
$RT_{C\!M\!O}$, respectively. 
Then the spurious equality ${RT}_{C\!O}~\peq~\mbox{min} ({RT}_C, {RT}_O)$ 
has the following three analogous generalizations:
\begin{eqnarray}
{RT}_{C\!M} &\peq &\mbox{min} ({RT}_C, {RT}_M), \\
{RT}_{M\!O} &\peq &\mbox{min} ({RT}_M, {RT}_O),  \quad \mbox{and} \\
{RT}_{C\!M\!O} &\peq & \mbox{min} ({RT}_C, {RT}_M, {RT}_O). 
\end{eqnarray}
Each of the above equalities should hold when V1 is assumed to have no neurons, i.e., the CM, MO, or even CMO neurons, which
are tuned to more than one feature dimension and can respond more vigorously to the corresponding double-feature 
(or triple-feature) singleton than it does to the corresponding singleton-feature singletons.

Fig. \ref{fig:OneSubject_Race5to8} shows the tests of all the four spurious equalities using behavioral 
data from the same observer SA used in Fig. \ref{fig:COSearch}.
In each case, the distribution of the reaction times of a multiple-feature singleton,
$RT_{C\!O}$,  
$RT_{M\!O}$,  
$RT_{C\!M}$,  
or $RT_{C\!M\!O}$,  is predicted from a race model involving the reaction times of the corresponding 
single-feature singletons. Each predicted distribution is significantly different from the observed 
distribution.

\subsection*{V1 neurons tuned conjunctively to color and orientation predict that ${RT}_{C\!O}$ is likely shorter than 
predicted by the race model}

Here we show that,  
because real V1 contains neurons (we call CO neurons) that are tuned simultaneously to 
color and orientation\cite{LivingstoneHubel1984},
the predicted $RT_{C\!O}$ using equality $RT_{C\!O}\peq {\rm min}(RT_C, RT_O)$ can be
longer than the observed  $RT_{C\!O}$.
Neurons tuned to color or orientation only are referred to as C or O neurons.
Iso-feature suppression implies that the
CO neuron responds more vigorously to a CO singleton
than to a background bar.  Let $r^{C\!O}$ denote the response of
the CO neuron, the maximum response to the CO singleton is then max$\left (r^C, r^O, r^{C\!O} \right )$, 
and according
to equation (\ref{eq:SingletonSaliency}), 
\begin{equation}
	{RT}_{C\!O} = f\left [{\rm max} \left (r^C, r^O, r^{C\!O} \right )\right ]. \label{eq:RTCOFromThreeResponses}
\end{equation}

Additionally, the CO neuron's response to the single-feature
C and O singletons are also likely higher than its response to a background bar. 
For example, a CO neuron tuned to green color and left tilt will respond to
a green, left-tilted bar when this bar is a C singleton (in a background of purple, left-tilted, bars),
or an O singleton (in a background of green, right-tilted, bars), or
an CO singleton (in a background of purple, right-tilted, bars).
The response level $r^{C\!O}$ is likely to distinguish between the three types of singletons,
since $r^{C\!O}$ is under iso-orientation suppression when the bar is a C singleton,
under iso-color suppression when the bar is an O singleton, and is free from iso-feature
suppression when the bar is a CO singleton.
To distinguish these responses, we use $r^{C\!O}_\alpha$ to denote the response of the
CO neuron to a singleton $\alpha = C$, $O$, or $CO$.
For completeness, we use $r^{C\!O}_B$ to denote the CO neuron's response to a 
background bar. Since $r^{C\!O}_B$ suffers from both iso-color and iso-orientation
suppression it is likely that  $r^{C\!O}_B<r^{C\!O}_\alpha$ for $\alpha = C$, $O$, and $C\!O$.

To be consistent and systematic, we similarly use $r^C_\alpha$ and $r^O_\alpha$ to denote
C and O neural response to a singleton bar  $\alpha= C$, $O$, and $C\!O$ or a background
bar $\alpha=B$.
For example, the responses of the C neuron to the four kinds of input bars are written as
$r^C_C$, $r^C_O$, $r^C_{C\!O}$, and  $r^C_B$.
We have previously ignored $r^C_O$ and identified $r^C_{C\!O}$ with $r^C_C$ since we argued that 
(ignoring response stochasticity for simplicity) 
\begin{equation}
r^C_B = r^C_O   \quad \mbox{with iso-color suppression and}\quad r^C_C = r^C_{C\!O} \quad \mbox{without iso-color suppression}, \label{eq:CBCO}
\end{equation}
because a C neuron response should only be affected by the presence of absence of iso-color 
suppression to make the orientation feature irrelevant. 
Similarly, the O neuron has the follow two distinct levels of responses,
\begin{equation}
r^O_B = r^O_C   \quad \mbox{and}\quad r^O_O = r^O_{C\!O}. \label{eq:OBOC}
\end{equation}
We will refer to neural responses such as $r^C_O (=r^C_B)$ and $r^O_C (=r^O_B)$ that can be 
equated with the same neurons' responses to a background bar as trivial responses.
 
Note that the meaning of, e.g., $C$, in a mathematical expression here
depends on whether it is a superscript or a subscript. As a superscript in, e.g., $r^C$ 
it means that the neuron giving the response is tuned to the color (C) feature; as 
a subscript in,  e.g., $r^O_C$  or ${RT}_C$ it means the visual input bar evoking the
response or reaction time is a color (C) singleton.
For simplicity and without loss of validity, we always ignore responses from neurons not 
tuned to the feature(s) of the bars, since their responses will always be smaller 
and will not affect the saliency values dictated by the maximum response to each location.

Combining equation (\ref{eq:SingletonSaliency}) with the equations above, we have
\begin{eqnarray}
{RT}_C &=& f\left [{\rm max}\left (r^C_C, r^O_C, r^{C\!O}_C\right )\right ] = f\left [{\rm max}\left (r^C_C, r^O_B, r^{C\!O}_C\right )\right ], \label{eq:RTc}\\
{RT}_O &=& f\left [{\rm max}\left (r^C_O, r^O_O, r^{C\!O}_O\right )\right ] = f\left [{\rm max}\left (r^C_B, r^O_O, r^{C\!O}_O\right )\right ], \label{eq:RTo}\\
{RT}_{C\!O} &=& f\left [{\rm max}\left (r^C_{C\!O}, r^O_{C\!O}, r^{C\!O}_{C\!O}\right )\right ] = f\left [{\rm max}\left (r^C_C, r^O_O, r^{C\!O}_{C\!O}\right )\right ]. \label{eq:RTco}
\end{eqnarray}
Since a C singleton bar is more salient than a background bar, 
 by V1 saliency hypothesis, its maximum evoked response
${\rm max}\left (r^C_C, r^O_B, r^{C\!O}_C \right )$ must be larger than the maximum 
response ${\rm max}\left (r^C_B, r^O_B, r^{C\!O}_B \right )$
evoked by a background bar, i.e., 
${\rm max}\left (r^C_C, r^O_B, r^{C\!O}_C \right ) > {\rm max}\left (r^C_B, r^O_B, r^{C\!O}_B \right )$.
Combining this with ${\rm max}\left (r^C_B, r^O_B, r^{C\!O}_B \right ) \ge r^O_B$ gives
 ${\rm max}\left (r^C_C, r^O_B, r^{C\!O}_C \right ) > r^O_B$, which in turn leads to ${\rm max}\left (r^C_C, r^O_B, r^{C\!O}_C \right )= {\rm max}\left (r^C_C, r^{C\!O}_C \right )$. 
Similarly ${\rm max} \left (r^C_B, r^O_O, r^{C\!O}_O\right )={\rm max}\left (r^O_O, r^{C\!O}_O \right )$.
Hence, we can ignore  $r^C_B$ and $r^O_B$ in equations (\ref{eq:RTc}--\ref{eq:RTo})  to have
\begin{equation}
{RT}_C = f\left [{\rm max}\left (r^C_C, r^{C\!O}_C\right )\right ] \quad \mbox{and}\quad 
{RT}_O = f\left [{\rm max}\left (r^O_O, r^{C\!O}_O\right )\right ]. \label{eq:RTcAndRTo}
\end{equation}
The above two equalities (compare them with equations (\ref{eq:RTc}) and (\ref{eq:RTo})) 
are just examples of the following equality for our singleton scenes:
\begin{equation}
\mbox{reaction time to a singleton}
= f[{\rm max}(\mbox{list of all non-trivial neural responses to this singleton})]. \label{eq:RT=ResponseList}
\end{equation}
This can be seen by reminding ourselves that a $r^O_C (= r^O_B)$ is a trivial response 
(i.e., statistically the same as the neuron's response to a background bar) to a C singleton whereas
$r^C_O (=r^C_B)$ is a trivial response to an O singleton.  
Continuing from equation (\ref{eq:RTcAndRTo}), 
\begin{eqnarray}	
	 {\rm min}({RT}_C, {RT}_O) &=&  {\rm min}\left \{
                        f\left [{\rm max}\left (r^C_C, r^{C\!O}_C \right ) \right ],
                        f\left [{\rm max}\left (r^O_O, r^{C\!O}_O \right )\right ] \right \}  \nonumber \\
&=&  f\{{\rm max}\left [{\rm max}\left (r^C_C, r^{C\!O}_C \right ), {\rm max}\left (r^O_O, r^{C\!O}_O \right )\right ] \} \nonumber \\
	&=& f\left [{\rm max}\left (r^C_C, r^O_O, r^{C\!O}_C, r^{C\!O}_O \right )\right ], \label{eq:RaceRTcAndRTo}
\end{eqnarray}
in which the second line arises from noting that $f(.)$ is a monotonically decreasing function,
the third line arises from the equality ${\rm  max}({\rm max}(a, b), {\rm max}(c, d), ...) = {\rm  max}(a, b, c, d, ...)$.
From equations (\ref{eq:RTcAndRTo}-\ref{eq:RaceRTcAndRTo}) above, we can see that 
equation (\ref{eq:RaceRTcAndRTo}) is a special case of the general equality
\begin{eqnarray}
{\rm min}(\mbox{list of reaction times for various singletons})&&  \nonumber \\
= f[{\rm max}(\mbox{list of all non-trivial}&&\!\!\!\!\!\!\!\!\!\! \mbox{neural responses to these singletons})]. \label{eq:RTlist=ResponseList}
\end{eqnarray}
This equality is the extension of equation (\ref{eq:RT=ResponseList})
to multiple (two or more) reaction times (for multiple singletons, each alone in one input scene), 
and holds for all our singleton scenes. It will be used to derive other race equalities.

Comparing ${\rm min}({RT}_C, {RT}_O) =f\left [{\rm max}\left (r^C_C, r^O_O, r^{C\!O}_C, r^{C\!O}_O \right )\right ]$ 
with $RT_{C\!O} = f\left [{\rm max}\left (r^C_C, r^O_O, r^{C\!O}_{C\!O} \right )\right ]$ 
(equation (\ref{eq:RTco})), we see that
${RT}_{C\!O} \peq {\rm min}({RT}_C, {RT}_O)$ requires
${\rm max}\left (r^C_C, r^O_O, r^{C\!O}_{C\!O} \right ) \peq {\rm max}\left (r^C_C, r^O_O, r^{C\!O}_C, r^{C\!O}_O \right )$.
This requirement can be met either by
\begin{equation}
{\rm max}\left (r^{C\!O}_C, r^{C\!O}_O, r^{C\!O}_{C\!O} \right ) < {\rm max}\left (r^C_C, r^O_O \right ),
\label{eq:NegligibleCO}
\end{equation}
which makes both ${\rm max}\left (r^C_C, r^O_O, r^{C\!O}_{C\!O} \right )$
and ${\rm max}\left (r^C_C, r^O_O, r^{C\!O}_C, r^{C\!O}_O \right )$
become simply ${\rm max}\left (r^C_C, r^O_O \right )$, 
or 
\begin{equation}
r^{C\!O}_{C\!O} ~\peq~ {\rm max}\left (r^{C\!O}_C, r^{C\!O}_O \right ),   \label{eq:COThreeResponses}
\end{equation}
which means that $r^{C\!O}_{C\!O}$ and $ {\rm max}\left (r^{C\!O}_C, r^{C\!O}_O \right )$ have the same distribution.
Note that inequality (\ref{eq:NegligibleCO}) can be 
satisfied when the CO responses $r^{C\!O}_C$, $r^{C\!O}_{O}$, and $r^{C\!O}_{C\!O}$
 are negligible relative to the C and O responses $r^C_C$ and $r^O_O$. 

The two conditions, equations (\ref{eq:NegligibleCO}) and (\ref{eq:COThreeResponses}),  can both be satisfied when CO neurons are absent so that 
$r^{C\!O}_C = r^{C\!O}_O = r^{C\!O}_{C\!O} = 0$. 
In this paper, a prediction 
e.g., a predicted equality such as ${RT}_{C\!O} \peq {\rm min}({RT}_C, {RT}_O)$, 
is called a spurious prediction 
if the neural properties (such as the two conditions above) upon which it relies are either known to be 
violated in V1, or whose presence in V1 is largely uncertain.
Whether the neural properties required for a spurious prediction can be satisfied or not may depend on individual observers, 
whose neural and behavioral sensitivities and feature selectivities are likely individually specific (e.g., some observers
may be color weaker than others).

Meanwhile, the race equality ${RT}_{C\!O} \peq  {\rm min}({RT}_C, {RT}_O)$ is likely broken when 
the CO neurons are present.  Iso-feature suppression makes it likely that
\begin{equation}
\left \la r^{C\!O}_{C\!O} \right \ra > \left \la {\rm max}\left (r^{C\!O}_C, r^{C\!O}_O \right ) \right \ra, 
\end{equation}
where $\la x \ra $ means the ensemble average of $x$.
If so, the equality ${RT}_{C\!O} ~ \peq~  {\rm min}({RT}_C, {RT}_O)$ 
is likely replaced by a race inequality
\begin{equation}
\la {RT}_{C\!O} \ra <  \left \la {\rm min}({RT}_C, {RT}_O) \right \ra . \label{eq:COraceViolation}
\end{equation}
Hence, the V1 saliency hypothesis makes the {\it qualitative} prediction that
${RT}_{C\!O}$ is likely to be statistically shorter than 
predicted by the race model; it cannot however predict quantitatively 
how much shorter this ${RT}_{C\!O}$ should be.
Meanwhile, breaking the equality ${RT}_{C\!O}~ \peq~ {\rm min}({RT}_C, {RT}_O)$ may also be 
manifested merely by a different distribution of ${RT}_{C\!O}$ from that of
${\rm min}({RT}_C, {RT}_O)$, rather than by a clear difference between their respective averages.

Similarly, V1 also contains MO neurons that are tuned simultaneously to orientation and motion 
direction\cite{HubelWieselFerrierLecture1977}. Hence, the following
inequality 
\begin{equation} \la {RT}_{M\!O}  \ra <  \la {\rm min}({RT}_M, {RT}_O) \ra, \label{eq:MOraceViolation}
\end{equation}
analogous to $\la {RT}_{C\!O} \ra <  \left \la {\rm min}({RT}_C, {RT}_O)\right \ra $, is likely to hold. 
However, V1 is reported to contain few CM neurons that are tuned simultaneously to color and motion direction\cite{HorwitzAlbright2005}, 
although conflicting reports\cite{HorwitzAlbright2005, Michael1978,TamuraEtAl1996} make it unclear
whether CM neurons are indeed absent or just fewer. Hence, it is unclear 
whether ${RT}_{C\!M} ~ \peq~  {\rm min}({RT}_C, {RT}_M)$
may be broken or whether the inequality $ \la {RT}_{C\!M}  \ra <  \la {\rm min}({RT}_C, {RT}_M) \ra$ may occur.

For observer SA in Fig. \ref{fig:OneSubject_Race5to8}, 
the behaviorally observed $\la {RT}_{C\!O} \ra$ and $\la {RT}_{M\!O}  \ra$ are indeed
shorter than their respective race model predicted 
values $\left \la {\rm min}({RT}_C, {RT}_O) \right \ra $ and
$\left \la {\rm min}({RT}_M, {RT}_O) \right \ra $, respectively. However, 
although ${RT}_{C\!M} \peq {\rm min}({RT}_C, {RT}_M)$ is violated, 
this observer has  $\la {RT}_{C\!M} \ra \approx \left \la {\rm min}({RT}_C, {RT}_M) \right \ra $.

The inequality
$\la {RT}_{\alpha\!\alpha '} \ra <  \left \la {\rm min}({RT}_\alpha, {RT}_{\alpha '}) \right \ra $  for $\alpha$ or $\alpha ' =C$, 
$M$, or $O$ and $\alpha \ne \alpha '$ is called a double-feature advantage or redundancy gain, and 
has been observed previously.  
Focusing on the time bins for the shortest reaction times, Krummenacher et al\cite{KrummenacherEtAl2001} 
showed that the density of ${RT}_{C\!O}$ in these bins were more than the summation of the
densities of the racers ${RT}_C$ and ${RT}_O$. Koene and Zhaoping\cite{KoeneZhaoping2007}
showed that
$\la {RT}_{C\!O} \ra < \left \la {\rm min}({RT}_C, {RT}_O) \right \ra$ 
and 
$\la {RT}_{M\!O} \ra < \left \la {\rm min}({RT}_M, {RT}_O) \right \ra$ 
hold statistically across the eight observers, 
whereas the average $\la {RT}_{C\!M} \ra$ is not significant different from
$\left \la {\rm min}({RT}_C, {RT}_M) \right \ra$. 
The current work extends the previous findings by comparing
the whole distribution of the observed ${RT}_{\alpha \!\alpha '}$
with its  race model prediction, i.e., the distribution of ${\rm min}({RT}_\alpha , {RT}_{\alpha '}) $.
The difference between the observed and the race-model predicted distributions should reflect
the contribution of the double-feature tuned neurons CO, MO, or CM, respectively, to the
saliency of the double-feature singleton (via its response $r^{C\!O}_{C\!O}$, $r^{M\!O}_{M\!O}$, or $r^{C\!M}_{C\!M}$, respectively, beyond the contribution of these neurons to the saliency of the 
single-feature singletons), as evaluated by Zhaoping and Zhe\cite{ZhaopingZhe2012}.

It is straightforward to generalize our derivations (in equations (\ref{eq:RTCOFromThreeResponses}--\ref{eq:COraceViolation})) 
to show that the spurious triple-feature race equality  $RT_{C\!M\!O}\peq {\rm min}(RT_C, RT_M, RT_O)$ is likely broken
when the responses from the double-feature tuned neurons are not negligible unless, analogous to 
equation (\ref{eq:COThreeResponses}), the response equality 
${\rm max}\left (r^{C\!M}_{C\!M}, r^{C\!O}_{C\!O}, r^{M\!O}_{M\!O} \right ) \peq
 {\rm max}\left (r^{C\!M}_C, r^{C\!M}_M, r^{C\!O}_C, r^{C\!O}_O, r^{M\!O}_M, r^{M\!O}_O \right )$ holds. Here, 
$r^{C\!M}_{\alpha}$ and 
$r^{M\!O}_{\alpha}$ are responses of the CM and MO neurons, respectively to single- or 
double-feature singleton $\alpha$, and we are assuming that V1 has no CMO cells tuned 
simultaneously to all three feature dimensions. 
Additionally,
just as $\la {RT}_{C\!O} \ra <\la {\min}({RT}_C, {RT}_O) \ra $ can result from 
$ \la r^{C\!O}_{C\!O} \ra > \left \la {\rm max}\left (r^{C\!O}_C, r^{C\!O}_O \right )  \right \ra $,
the race inequality $\la {RT}_{C\!M\!O} \ra < \la {\rm min}({RT}_C, {RT}_M, {RT}_O) \ra $ can result from 
the neural response inequality 
\begin{equation}
\left \la {\rm max}\left (r^{C\!M}_{C\!M}, r^{C\!O}_{C\!O}, r^{M\!O}_{M\!O} \right )  \right \ra >
 \left \la {\rm max}\left (r^{C\!M}_C, r^{C\!M}_M, r^{C\!O}_C, r^{C\!O}_O, r^{M\!O}_M, r^{M\!O}_O \right )  \right \ra,  
\end{equation}
which can arise when the double-feature tuned neurons respond more vigorously to the double- or triple-feature 
singletons than to the single-feature singletons due to iso-feature suppression.

The above inequality is like a composite of the three component inequalities
$\la r^{C\!O}_{C\!O}\ra > \la {\rm max}(r^{C\!O}_C, r^{C\!O}_O) \ra $, 
$\la r^{M\!O}_{M\!O} \ra > \la {\rm max}(r^{M\!O}_M, r^{M\!O}_O) \ra$, and 
$\la  r^{C\!M}_{C\!M} \ra > \la {\rm max}(r^{C\!M}_C, r^{C\!M}_M) \ra $.  Hence,
it is likely to hold when two out of the three component inequalities hold.
According to analysis around equations (\ref{eq:COThreeResponses}--\ref{eq:COraceViolation}), 
each component inequality 
$\la r^{\alpha\!\alpha '}_{\alpha\!\alpha '} \ra > 
\la {\rm max}(r^{\alpha\!\alpha '}_\alpha, r^{\alpha\!\alpha '}_{\alpha '}) \ra $
is implied by the corresponding race inequality
$\la {RT}_{\alpha\!\alpha '}\ra < \la {\rm min}({RT}_\alpha, {RT}_{\alpha '}) \ra $ for
${\alpha\!\alpha '} = CO$, $MO$, or $CM$.
Therefore, the triple-racer inequality $\la {RT}_{C\!M\!O} \ra < \la {\rm min}({RT}_C, {RT}_M, {RT}_O) \ra $ 
is quite likely when two out of the three double-racer inequalities 
$\la {RT}_{\alpha\!\alpha '}\ra < \la {\rm min}({RT}_\alpha, {RT}_{\alpha '}) \ra $ 
hold. 
This is the case for the observer's data in Fig. \ref{fig:OneSubject_Race5to8}.
Meanwhile, we note that the composite equality
${\rm max}\left (r^{C\!M}_{C\!M}, r^{C\!O}_{C\!O}, r^{M\!O}_{M\!O} \right ) \peq
 {\rm max}\left (r^{C\!M}_C, r^{C\!M}_M, r^{C\!O}_C, r^{C\!O}_O, r^{M\!O}_M, r^{M\!O}_O \right )$
does not necessarily break when the component equality 
$r^{\alpha\!\alpha'}_{\alpha\!\alpha'} \peq {\rm max}\left (r^{\alpha\!\alpha'}_\alpha, r^{\alpha\!\alpha'}_{\alpha'}\right )$ is broken for each $\alpha\!\alpha' = C\!O$, $M\!O$, and $C\!M$ (just as
equality ${\rm max}(5, 4, 6) ={\rm max}(3, 4, 2, 6, 4, 5)$ holds despite
$5 \ne {\rm max}(3, 4)$, $4 \ne {\rm max}(2, 6)$, and $6 \ne {\rm max}(4, 5)$).

\subsection*{A quantitative prediction of the reaction time for a triple-feature singleton from another race equality}

We have seen that
the presence of the CO neurons likely breaks  ${RT}_{C\!O} ~ \peq ~ {\rm min}({RT}_C, {RT}_O)$,
making ${RT}_{C\!O}$ not predictable quantitatively from ${RT}_C$ and ${RT}_O$ even though one may 
qualitatively expect $\la {RT}_{C\!O}\ra  < \la {\rm min}({RT}_C, {RT}_O) \ra$ as likely. 

To make a quantitative prediction, we need to go further by finding a type of
joint feature selectivity that at most barely exists in V1 neurons. 
This motivates us to consider CMO neurons tuned simultaneously to all the three features, C, M, and O.
Given the existing paucity of V1 neurons tuned simultaneously to C and M\cite{HorwitzAlbright2005}, 
we can be far more confident that CMO cells (which should at least be tuned simultaneously to C and M) 
are absent in V1. Just as the absence of CO neurons gives ${RT}_{C\!O} ~ \peq ~ {\rm min}({RT}_C, {RT}_O)$,
the absence of the CMO neurons gives this race equality 
\begin{equation}
 \mbox{min}({RT}_{C\!M\!O}, {RT}_C, {RT}_M, {RT}_O) \peq \mbox{min} ({RT}_{C\!M}, {RT}_{C\!O}, {RT}_{M\!O}),   
 \label{eq:CMOrace}
\end{equation}
see Methods for its proof.
The left side of the equality is the race outcome from four racers with their respective reaction times as
${RT}_{C\!M\!O}$, ${RT}_C$, ${RT}_M$, and ${RT}_O$,
and the right side is the race outcome of another three racers with their respective reaction times as
${RT}_{C\!M}$, ${RT}_{C\!O}$, and ${RT}_{M\!O}$. 
This race equality thus states that the two different races produce the same distribution of winning reaction times.
Since we are quite confident about the condition (that CMO cells are absent in V1) behind this equality,
we call this a non-spurious race equality. 
It enables us to predict the distribution of ${RT}_{C\!M\!O}$ from those of the other six types of reaction times.
We call this prediction our non-spurious prediction, which can be compared with behaviorial data as shown next.

\subsection*{The non-spurious race equality holds across all six observers}

\begin{figure}[hhhhhhthh!!]
\begin{center}
\includegraphics[width=120mm]{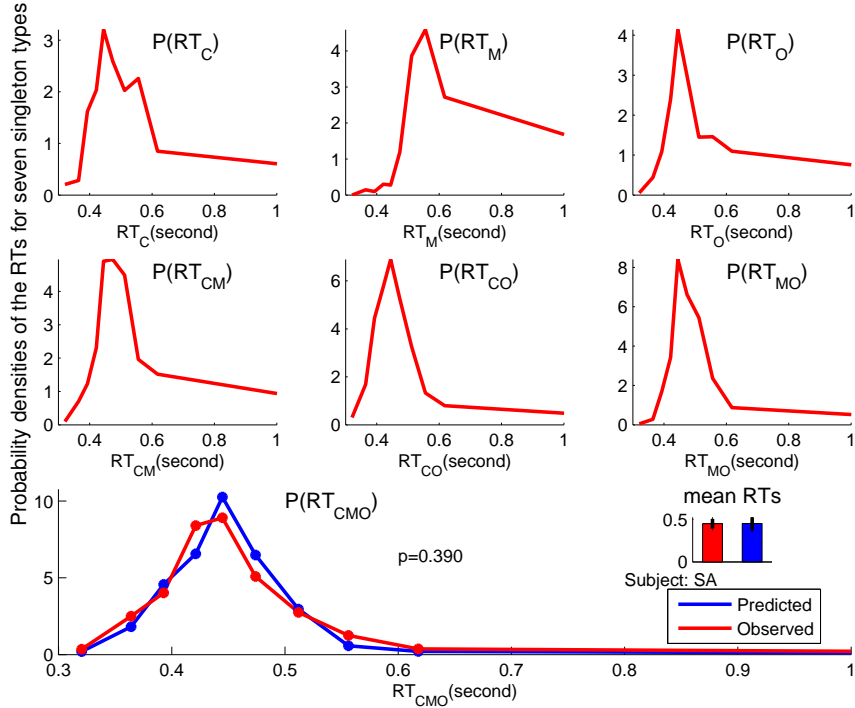}
\end{center}
\caption{ \label{fig:CMOSearch}
The observed distributions of ${RT}_C$, ${RT}_M$, ${RT}_O$, ${RT}_{C\!M}$, ${RT}_{C\!O}$, and ${RT}_{M\!O}$ for an observer
are used to predict the distribution of ${RT}_{C\!M\!O}$ for the same observer (SA, the same as that in
Figs. \ref{fig:COSearch} and \ref{fig:OneSubject_Race5to8}) by the non-spurious race equality 
$\mbox{min}({RT}_{C\!M\!O}, {RT}_C, {RT}_M, {RT}_O) \peq \mbox{min} ({RT}_{C\!M}, {RT}_{C\!O}, {RT}_{M\!O})$.
The predicted and observed distributions of
${RT}_{C\!M\!O}$  are statistically indistinguishable from each other  ($p = 0.39$).
This figure has the same format as Fig. \ref{fig:COSearch}.
There were 306, 296, 296, 308, 308, 311, and 312 behavioral data samples for 
${RT}_C$, 
${RT}_M$, 
${RT}_O$, 
${RT}_{C\!M}$, 
${RT}_{C\!O}$, 
${RT}_{M\!O}$, 
and ${RT}_{C\!M\!O}$. 
}
\end{figure}

\begin{figure}[hhhhhhthh!!]
\begin{center}
\includegraphics[width=120mm]{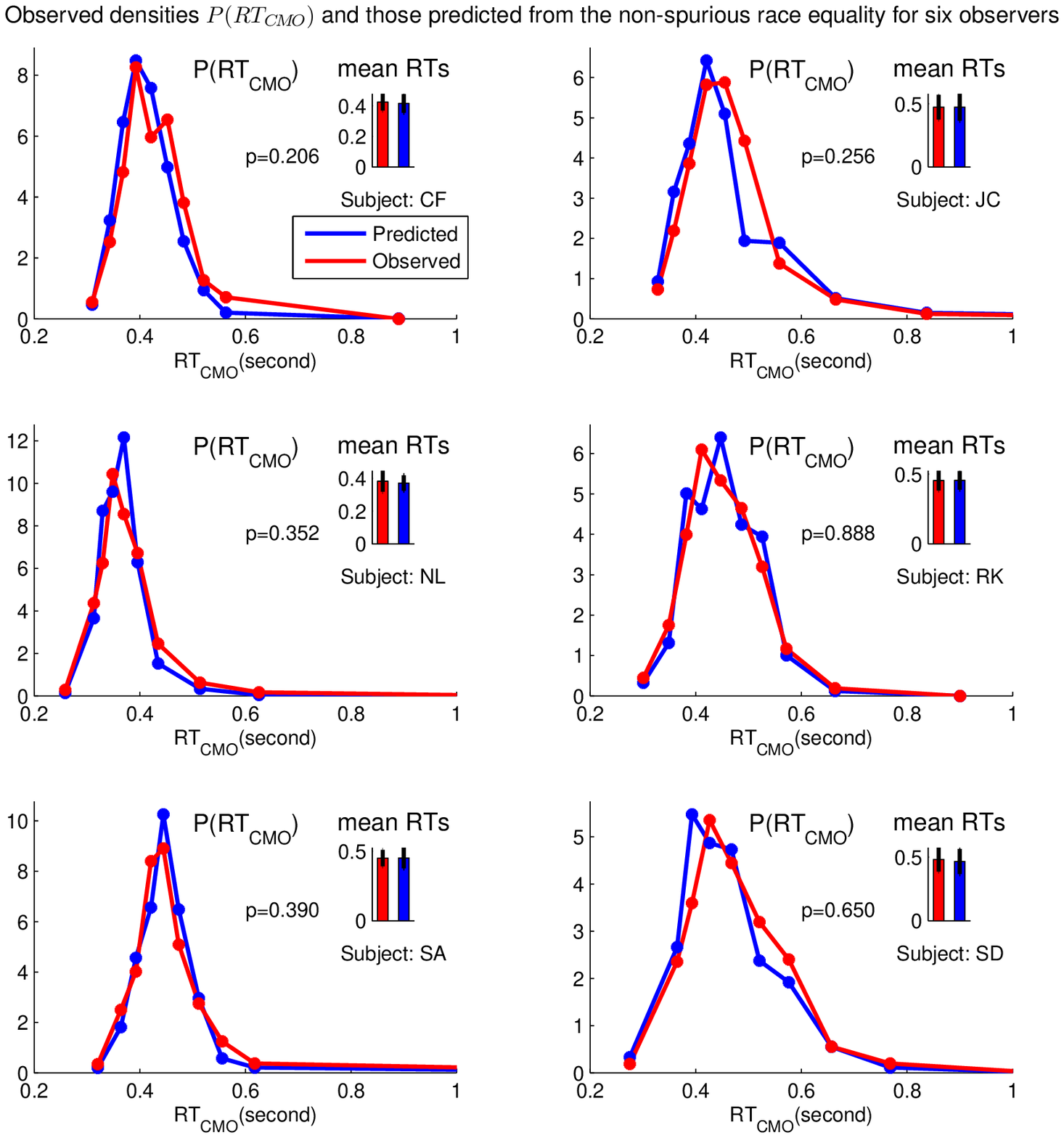}
\end{center}
\caption{ \label{fig:CMOSearch_AllObservers}
Observed and predicted distributions of ${RT}_{C\!M\!O}$ using the non-spurious race equality for  six observers, 
including observer SA whose details are shown in Fig. \ref{fig:CMOSearch}.
The predictions agree with data for all observers, indicated by the $p>0.05$.
}
\end{figure}

Fig \ref{fig:CMOSearch} shows that the observed distribution of ${RT}_{C\!M\!O}$ for our example observer SA
is statistically indistinguishable from the one predicted from the reaction times for the other six types of 
singletons using our non-spurious race equality.
Fig \ref{fig:CMOSearch_AllObservers} shows that this agreement between the predicted and the observed
${RT}_{C\!M\!O}$ holds for all six naive adult observers.

One may ask whether our non-spurious equality (equation (\ref{eq:CMOrace})) is
hard to falsify because it has a different and more complex structure than our spurious equalities
$RT_{\alpha\!\alpha '}\peq {\rm min}\left (RT_\alpha, RT_{\alpha '}\right )$ and
$RT_{C\!M\!O}\peq {\rm min}\left (RT_C, RT_M, RT_O \right )$. 
In each of the spurious race equalities, the reaction time to be predicted and the other
reaction times are on opposite sides of the equality.
In our non-spurious equality (equation (\ref{eq:CMOrace})), the $RT_{C\!M\!O}$ to be predicted
has to race with some other types of reaction times to contribute to the equality, 
making its prediction more complex (see Methods).  To show that this complexity in the race equality 
does not prevent a falsification of a spurious equality, 
we create three new spurious equalities that have
the same complexity as our non-spurious equality but can be falsified by our behavioral data.
Listing our non-spurious equality with these three newly created spurious equalities next to each other,
\begin{eqnarray}
\mbox{non-spurious}: & ~~~ {\rm min}({RT}_{C\!M\!O}, RT_C, RT_M, {RT}_O) & \peq~~~  {\rm min}({RT}_{C\!M}, RT_{C\!O}, {RT}_{M\!O}), \label{eq:RE1} \\
\mbox{spurious}: & ~~~ {\rm min}({RT}_{C\!M\!O}, {RT}_M, {RT}_{C\!O})& \peq~~~  {\rm min}({RT}_C, {RT}_O, {RT}_{C\!M}, {RT}_{M\!O}), \label{eq:RE6} \\
\mbox{spurious}: & ~~~ {\rm min}({RT}_{C\!M\!O}, {RT}_C, {RT}_{M\!O})& \peq~~~  {\rm min}({RT}_M, {RT}_O, {RT}_{C\!M}, {RT}_{C\!O}), \label{eq:RE7} \\
\mbox{spurious}: & ~~~ {\rm min}({RT}_{C\!M\!O}, {RT}_O, {RT}_{C\!M})& \peq~~~  {\rm min}({RT}_C, {RT}_M, {RT}_{C\!O}, {RT}_{M\!O}), \label{eq:RE8}
\end{eqnarray}
we can examine their similarities and relationships.
For example, between  the non-spurious equality and the first spurious one above,
their left sides of the equalities are identical to each other if
$RT_{C\!O}\peq {\rm min} (RT_C, RT_O)$ holds, so are their right sides of the equalities.
This means, the first spurious equality above is spurious when $RT_{C\!O}\peq {\rm min} (RT_C, RT_O)$ 
is spurious, unless $RT_C$, $RT_O$, and $RT_{C\!O}$ are likely losers in their respective races 
(${\rm min}({RT}_{C\!M\!O}, RT_C, RT_M, {RT}_O)$ and ${\rm min}({RT}_{C\!M}, RT_{C\!O}, {RT}_{M\!O})$)
in the 
non-spurious equality so that they do not matter.
Similarly, the second or third spurious equality is spurious when
 $RT_{M\!O}\peq {\rm min} (RT_M, RT_O)$ or
 $RT_{C\!M}\peq {\rm min} (RT_C, RT_M)$, respectively, is spurious, unless
the corresponding racers are likely losers in the non-spurious races. 
In other words, each of the three spurious equalities above is a corollary
of one of our previous, double-feature, spurious, race equalities 
$RT_{\alpha\!\alpha '}\peq {\rm min}\left (RT_\alpha, RT_{\alpha '}\right )$.
For convenience, we sometimes refer to 
$RT_{\alpha\!\alpha '}\peq {\rm min}\left (RT_\alpha, RT_{\alpha '}\right )$ as
the original spurious equalities to their respective corollary equalities.

\begin{figure}[hhhhhhthh!!]
\begin{center}
\includegraphics[width= 120mm]{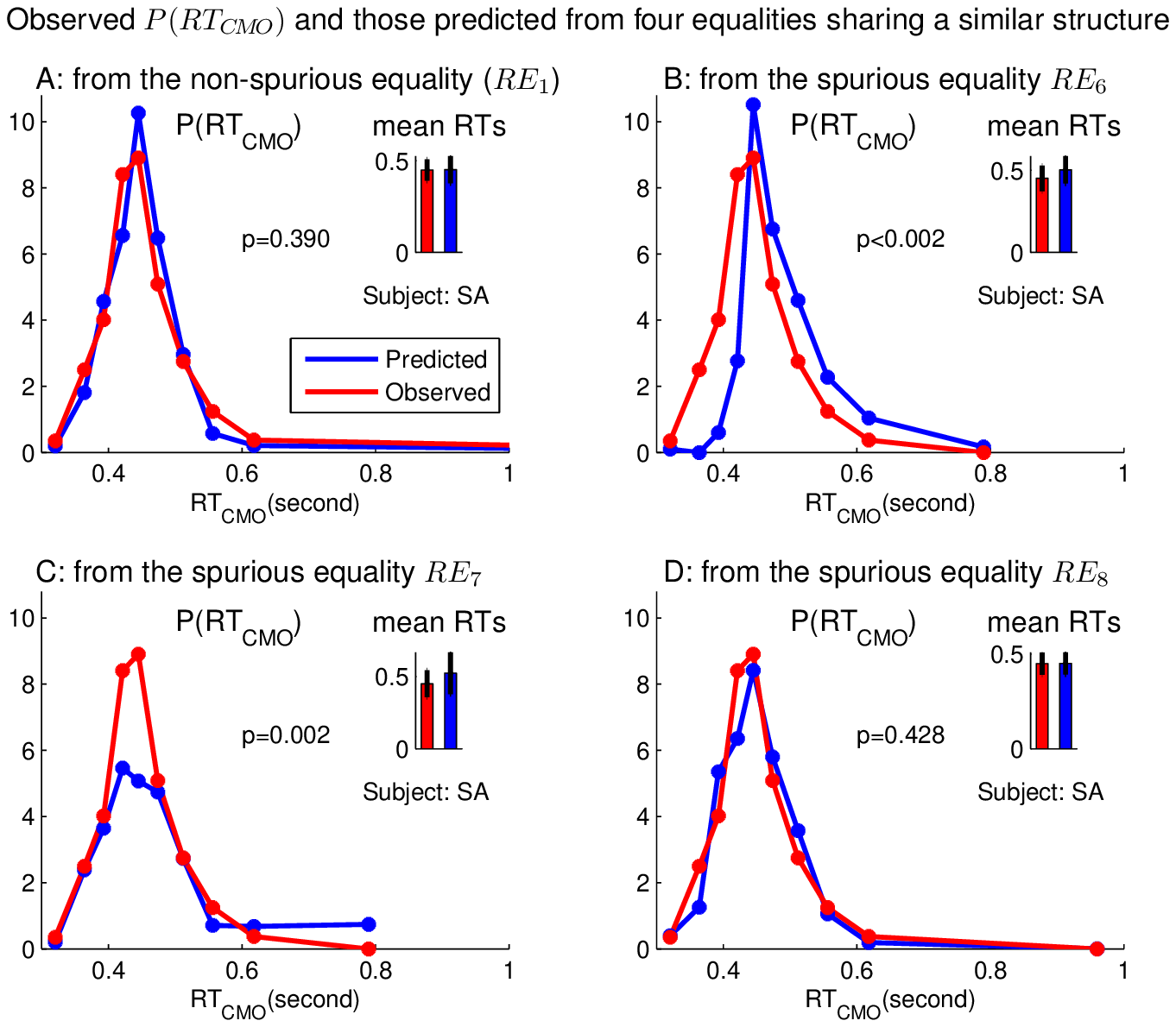}
\end{center}
\caption{ \label{fig:OneSubject_Race1to4}
The predicted and observed $P({RT}_{C\!M\!O})$ from the non-spurious equality and
three spurious ones sharing similar complexities, listed in equations (\ref{eq:RE1}--\ref{eq:RE8}). 
These equalities are also denoted by $RE_1$, $RE_6$, $RE_7$, and $RE_8$ in Table 1.
}
\end{figure}

Each of the four equalities above (one non-spurious) can be used to predict the distribution of $RT_{C\!M\!O}$ 
from those of the other six types of reaction times. Match or mismatch between the predicted and observed 
distributions of $RT_{C\!M\!O}$ indicates a confirmation or falsification, respectively, of the race equality.
Fig. \ref{fig:OneSubject_Race1to4} shows that,  in our example observer SA, the first two but not the last one of
the spurious, corollary, equalities above are falsified, whereas  Fig. \ref{fig:OneSubject_Race5to8}
has shown that all three of the original spurious equalities for this observer are falsified.
The lack of a falsification of the last corollary equality, despite the falsification 
of its original, can be comprehended as follows.
From the analysis in the last paragraph, the falsification of the first two corollary equalities
indicates that the corresponding reaction times, especially $RT_{C\!O}$ and $RT_{M\!O}$,    
are likely winners in the race of the non-spurious equality, making
the reaction time  $RT_{C\!M}$ likely a loser so that the violation
of the original equality $RT_{C\!M} \peq {\rm min}\left (RT_C, RT_M \right )$ is less likely
to be influential (i.e., to break the corollary equality).
Furthermore,  Fig. \ref{fig:OneSubject_Race5to8}C shows that the densities of the 
observed and the predicted ${RT}_{C\!M}$  from the original equality ${RT}_{C\!M}\peq {\rm min}\left (RT_C, RT_M \right )$
match well for the shortest reaction times, implying that the original equality is not
violated for the shortest reaction times, which are most likely to be race winners 
(for the non-spurious equality) to be influential.
Among all the six observers,  each corollary spurious equality is broken
in fewer observers than its original spurious equality.
Nevertheless, the breakings of the corollary spurious equalities, which are just as complex
as our non-spurious equality, demonstrate that complexity of a race equality is insufficient 
to prevent a  falsification.

\subsection*{Qualitative conclusions  from our reaction time data despite sensitivity of some findings to parameter variations in our data analysis method}

So far, we only illustrated the tests of all the spurious equalities using results 
from one observer, only the test of the non-spurious equality is presented
by results from all the six observers from whom we had collected reaction time data on all the 
seven types of singletons (see Methods).
Furthermore, all the tests have so far been illustrated 
using a particular set of parameters characterizing the technical details
in our precedures (see Methods) to test the race equalities.
We found that the qualitative conclusions of our study do not depend on these technical details. 
These details are characterized by four parameters (see Methods section): (1) 
the number $N$ of time bins used to discretize about 300 reaction time data samples for each 
singleton type of each observer,
(2) the way to determine the boundaries between the time bins given $N$, 
(3) the metric used to measure the distance $D$ between
the predicted and the behaviorally observed distributions of the reaction times to 
judge whether a race equality holds, 
and 
(4) (only applicable to the four more complex race equalities 
in equations (\ref{eq:RE1}--\ref{eq:RE8})),  
the objective metric, i.e., the distance between the distributions on the two sides of a 
race equality, to be minimized in the optimization procedure to predict 
the ${RT}_{C\!M\!O}$ distribution.
The example results in Figs. \ref{fig:COSearch}--\ref{fig:OneSubject_Race1to4} 
are obtained using the following set of 
parameters: (1) $N = 10$ (from one of five choices $N=8, 9, 10, 11, 12$), 
(2) reaction time bins are chosen using equation (\ref{eq:ChooseXF}) with
$x_F = 1.35$ (from four different choices listed around equation (\ref{eq:ChooseXF})),  
(3) the $D$ metric and (4) the objective metric are both chosen as the Hellinger distance 
(each is from one of the four choices, see equation (\ref{eq:FourDistances})). 
In this section, some general statistics of our findings across 
$5\times 4\times 4 = 80$ (or $5\times 4\times 4\times 4 = 320$ for the more complex equalities)
variations of the technical parameters are presented. 
In particular, we show the number of observers whose data break each spurious 
or non-spurious race equality, averaged across the variations of the technical parameters 
in the testing method. 

For convenience, Table 1 lists all the (spurious or non-spurious) race equalities, 
each is written in the format of $RT1 \peq RT2$ with a definition of $RT1$ and $RT2$. 
For example, the equality $RT_{C\!O}\peq {\rm min}(RT_C, RT_O)$ has
$RT1\equiv RT_{C\!O}$ and $RT2 \equiv  {\rm min}(RT_C, RT_O)$.
Each race equality (RE) is denoted by a race equality index $REI = 1$, $2$, ..., or $8$ so that
it will be referred to as $RE_1$, $RE_2$, ...or $RE_8$, respectively, for easy reference.
The $RE_1$, i.e., race equality $RE_i$ with $i=1$, is our (only) non-spurious equality 
${\rm min}({RT}_{C\!M\!O}, {RT}_C, {RT}_M, {RT}_O) \peq {\rm min}({RT}_{C\!M}, {RT}_{C\!O}, {RT}_{M\!O})$. 
The $RE_i$ for $i=2$--$4$ are the double-racer-model equalities
$RT_{\alpha\!\alpha '} \peq {\rm min}\left (RT_\alpha, RT_{\alpha '} \right )$ for
${\alpha\!\alpha '} = {C\!O}$, 
${M\!O}$, and ${C\!M}$, respectively.
The $RE_i$ for $i=6$--$8$ are their respective corollary (complex) equalities.
The $RE_5$ is the triple-racer-model equality
${RT}_{C\!M\!O}\peq {\rm min}\left ({RT}_C, {RT}_M, {RT}_O \right )$. 
For  each equality, one of the reaction times involved is designated
as the one whose distribution will be predicted from those of the other reaction times
using the equality. This designated reaction time
is named as $RT_{\rm goal}$ in Table 1. It
is always the one for the singleton with the largest number of unique features, 
thus it tends to be the shortest reaction time and thus is more precisely determined,
by the nature of the race(s), from the other reaction times involved in the race 
equality. Hence $RT_{C\!M\!O}$ is the $RT_{\rm goal}$  for all race equalities except
$RE_i$ with $i=2$--$4$, whose $RT_{\rm goal}$ are $RT_{\alpha\!\alpha '}$ 
for ${\alpha\!\alpha '} = {C\!O}$, ${M\!O}$, and ${C\!M}$, respectively. 

\centerline{Table 1: race equalities $RT1 ~\peq ~ RT2$ considered in this paper}

\vspace{0.1 in}
\begin{tabular}{cccc}
Equality &       $RT1$     &    $RT2$  &    {${RT}_{\rm goal}$ designated } \\
Type/label         &       &         &                          {for prediction } \\
\hline
Non-spurious & && \\
$RE_1$ & min(${RT}_{C\!M\!O}, {RT}_C, {RT}_M, {RT}_O$) & min (${RT}_{C\!M}, {RT}_{C\!O}, {RT}_{M\!O}$) & ${RT}_{C\!M\!O}$ \\
\hline
Spurious & && \\
$RE_2$ &  ${RT}_{C\!O}$  & min (${RT}_C, {RT}_O$) & ${RT}_{C\!O}$ \\
$RE_3$ &  ${RT}_{M\!O}$  & min (${RT}_M, {RT}_O$) & ${RT}_{M\!O}$ \\
$RE_4$ &  ${RT}_{C\!M}$  & min (${RT}_C, {RT}_M$) & ${RT}_{C\!M}$  \\
$RE_5$ &  ${RT}_{C\!M\!O}$ & min (${RT}_C, {RT}_M, {RT}_O$) & ${RT}_{C\!M\!O}$ \\
$RE_6$ & min(${RT}_{C\!M\!O}, {RT}_M, {RT}_{C\!O}$) & min (${RT}_C, {RT}_O, {RT}_{C\!M}, {RT}_{M\!O}$) & ${RT}_{C\!M\!O}$ \\
$RE_7$ & min(${RT}_{C\!M\!O}, {RT}_C, {RT}_{M\!O}$) & min (${RT}_M, {RT}_O, {RT}_{C\!M}, {RT}_{C\!O}$) & ${RT}_{C\!M\!O}$ \\
$RE_8$ & min(${RT}_{C\!M\!O}, {RT}_O, {RT}_{C\!M}$) & min (${RT}_C, {RT}_M, {RT}_{C\!O}, {RT}_{M\!O}$) & ${RT}_{C\!M\!O}$ \\
\hline
\end{tabular}

\vspace{0.3 in}

Koene and Zhaoping\cite{KoeneZhaoping2007} collected reaction time data 
for each of the single- and double-feature singletons
from eight observers, but collected $RT_{C\!M\!O}$ data from only six of these observers.
Hence, $RE_i$ with $i=2$--$4$ can be tested by eight observers while the other
equalities by only six observers.

Whether a race equality can be falsified by data from a particular observer depends on 
several factors. First, as mentioned before, it may depend on the observer, as there 
may be inter-observer difference in terms of the V1 properties and visual sensitivities. 
For example, some observers may be more color weak than others.
Second, even when a race equality is truely false for a particular observer, 
it may appear to hold from this observer's behavioral data when there are not enough 
samples of reaction time data to achieve a sufficient statistical power for revealing 
a difference between prediction and observation (especially when the deviation from a race equality 
is small). 
Meanwhile, even when a race equality is fundamentally true, 
there is a 5\% chance to find it accidentally broken by behavioral data.
This is because, by definition (see Methods), a null hypothesis proclaiming the race equality 
is declared as false when the distance $D$ between the predicted (by the race equality) 
and observed distributions of reaction times  is larger than 95\% of the random samples of the 
distances $D$ in the situation when the null hypothesis strictly holds. 
Third, empirically, we observed that in some occasions the technical parameters in our procedure 
can also affect whether a race equality is falsified by data.

\begin{figure}[hhhhhhthh!!]
\begin{center}
\includegraphics[width = 150mm]{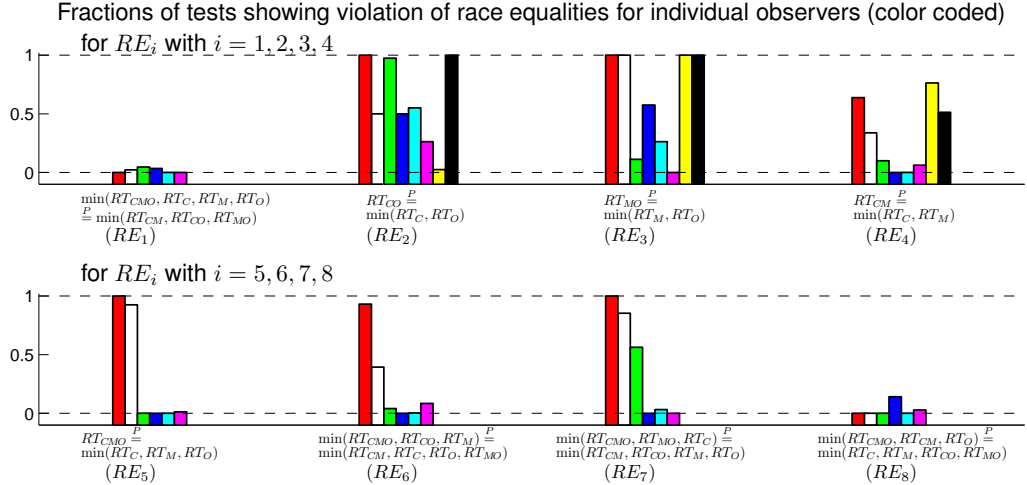}
\end{center}
\caption{ \label{fig:FractionOfViolations}
The fraction of the tests of each race equality in which the equality is falsified for each observer.
Each observer is color coded by a unique color: red, white, green, blue, cyan, magenta, yellow, or black (our example observer SA is coded by red).
Different tests of a race equality use different sets of parameters in the testing method to
include all possible combinations of the parameter values specified in the Methods section.
Each race equality is tested on six observers except for $RE_i$ for $i=2$--$4$ which 
include two additional observers coded by yellow and black data bars, respectively.
Each equality  $RE_i$  for $i=2$--$4$ and its corollary equality $RE_{i+4}$ are positioned in a 
vertically aligned manner for easy of comparison.
}
\end{figure}

Given a race equality and an observer, if parameter variations for the tests do not sensitively 
affect the qualitative outcome of the test, then the fraction of all the ($80$ or $320$) 
tests in which the equality is found broken should be close to $1$ or $0$ to indicate that the race equality 
is consistently broken or maintained, respectively.
Fig. \ref{fig:FractionOfViolations} plots these fractions across observers and race equalities.
Among 54 different combinations of observers and race equalities, 34 
give this fraction as either larger than 95\% or smaller than 5\%, and 11 have this fraction
closer to $50\%$ than to either 100\% or 0\%.
Sensitivity of the test to the test parameters are mainly caused by 
the sensitivity to the metric used to  measure the difference between the predicted 
and observed distributions of reaction times.  
We found that for some observers in some race equalities, e.g., 
observers marked by white, blue, and cyan color for $RE_2$,
a race equality is consistently broken using one metric and consistently maintained
using another metric, (almost) regardless of the variations of the other parameters for the tests.

\begin{figure}
\begin{center}
\includegraphics[width=140mm]{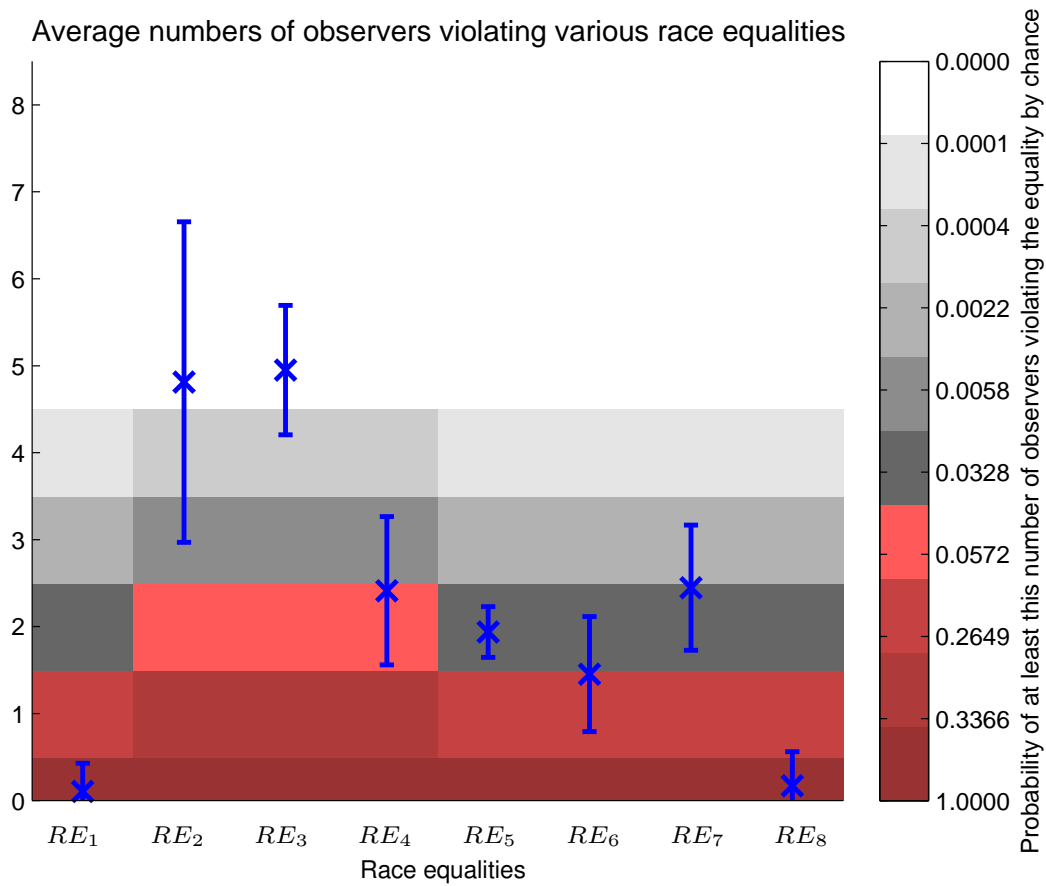}
\end{center}
\caption{\label{fig:NumberOfObserversBreakEquality}
Average numbers of observers to break various race equalities, as shown in blue data points whose 
error bars denote standard deviations. The non-spurious race equality is $RE_1$.
Data from $6$ observers were tested for race equalities $RE_1$ and $RE_i$ for $i\ge 5$ and data from 8 observers were tested 
for $RE_i$ for $i=2$--$4$. 
Applying a test of a given race equality to all the observers gives a number of observers breaking this equality, 
and the average of this number over $80$ (for $RE_i$ with $i=2$--$5$) or  $320$ (for $RE_1$ and $RE_i$ with $i>5$) tests, 
each  characterized by a unique set of parameters in the testing method, 
gives the blue data point for this equality.
The background shadings visualize the probabilities of at least a certain number 
of observers breaking a true race equality accidentally, those probabilities larger than $0.05$ are 
visualized by shades with a red hue. 
Note that the number of observers in this probability representation is an integer number, 
whereas the data points are generally non-integers since they are averages of integer numbers.
}
\end{figure}

Since each observer has a 5\% chance 
to accidentally break a true race equality,
one expects that, among $N=6$ or $8$ observers, an average of 
$0.05\cdot N=0.3$ or $0.4$ observers, respectively, to break a true race equality accidentally.
More generally, there is a chance of ${N! \over {n!(N-n)!}}0.05^n 0.95^{N-n}$ that 
$n$ out of $N$ observers will break this true race equality accidentally.
Accordingly, for six observers, there is a chance probability of  27\%,  3\%, or 0.2\%, respectively, that 
at least one, two, or three observers accidentally break a true race equality; for eight observers,
the corresponding chance becomes  34\%, 6\%, or 0.6\% respectively.
Therefore, if more than one or two out of six or eight observers, respectively, break a race equality, we say that
the equality is broken or incorrect since such a high tendency of equality breaking can happen only by a chance of 
less than 0.05 for a truely correct race equality.

Individual differences in neural response properties and a lack of statistical power in data are 
likely to partly explain why even the most obviously spurious equality ${RT}_{C\!O} \peq {\rm min}({RT}_C, {RT}_O)$ is not broken by data from all observers.  
For example, the observer coded by yellow color in Fig. \ref{fig:FractionOfViolations} appears to 
show race equality ${RT}_{C\!O} \peq {\rm min}({RT}_C, {RT}_O)$; this may either be caused by a 
lack of vigorously responding CO cells in this observer, or it may be because the difference 
between ${RT}_{C\!O}$ and ${\rm min}({RT}_C, {RT}_O)$ is too small to be detected by around 300 
random samples of each type of reacdtion times ${RT}_{C\!O}$, ${RT}_C$, and ${RT}_O$.

Fig. \ref{fig:NumberOfObserversBreakEquality} plots the numbers of our observers to break various
race equalities.  Each number is the average over the outcomes of all the tests (each applied
to all individual observers) of a race equality using different sets of parameters in the testing method.
Data points on gray or white background are those with more observers
breaking an equality than expected by a probability of $0.05$ if the equality truely holds.
These results enable the following qualitative conclusions which are relatively immune to the sensitivities 
to detailed parameters in the testing method.  First, the non-spurious 
race equality ($RE_1$) is confirmed by our data since it is only broken by an average of $0.1$ observers, within the range expected 
for chance breaking of a true equality.
Second, two spurious predictions, $RE_2$ and $RE_3$ (for equalities 
${RT}_{C\!O} \peq {\rm min}({RT}_C, {RT}_O)$ and  ${RT}_{M\!O} \peq {\rm min}({RT}_M, {RT}_O)$, respectively), are broken since 
data from more than an average of $4$ observers break each of them.
This finding is consistent with physiology that there are CO and MO neurons in V1\cite{HubelWieselFerrierLecture1977, LivingstoneHubel1984}.
Third, the spurious prediction $RE_4$ for equality 
${RT}_{C\!M} \peq {\rm min}({RT}_C, {RT}_M)$ is marginally broken, or not as seriously broken as 
the spurious predictions $RE_2$ and $RE_3$,  since it is broken by an average of only $2.4$ out of 
eight observers. This is consistent with the idea that V1 has fewer CM neurons compared 
to CO and MO neurons,  and is consistent with the controversy in experimental 
reports\cite{HorwitzAlbright2005, Michael1978,TamuraEtAl1996}
regarding the presence or absence of the CM cells.
Fourth, the spurious prediction $RE_5$ for equality ${RT}_{C\!M\!O} \peq {\rm min}({RT}_C, {RT}_M, {RT}_O)$ is broken since data
from an average of $1.94$ out of six observers violate it.
This is consistent with the fact that V1 contains a substantial number of conjunctively tuned cells, 
in particular the CO and MO cells, and corroborates the finding that
its component race equalities $RE_2$ and $RE_3$ are clearly broken.  
Fifth, the more complex spurious equalities $RE_{i}$ for $i=6$--$8$,  each a variation of 
the non-spurious $RE_1$ and can be potentially undermined 
(when certain conditions hold, as discussed in 
the text around equations (\ref{eq:RE1}--\ref{eq:RE8})) by the violation of the
 corresponding original $RE_{2-4}$, are marginally broken, broken, and maintained, respectively, 
in our data. 
This corroborates our findings for the original spurious $RE_{2-4}$.
Additionally, weaker violations of the corollary equalities 
compared to the degrees of violations in their original counterparts lend further support 
to our non-spurious $RE_1$, which contributes to sustaining the corollary equalities against the 
undermining factors from the violations of the original equalities.

We note that our non-spurious $RE_1$ and the spurious $RE_i$ for $i=6$--$8$ have very similar 
structures and use the same technical procedure to predict ${RT}_{C\!M\!O}$ from the same set of 
other types of reaction times. Hence, a clear rejection of race equality $RE_7$
by our data indicates that our data have a sufficient statistical power to reject 
our non-spurious equality $RE_1$ if it were as clearly incorrect as $RE_7$.
Therefore, our non-spurious V1 prediction is confirmed  at least within the resolution 
provided by the statistical power of our data. This resolution is manifested 
in Fig. \ref{fig:OneSubject_Race1to4} in which it can
clearly distinguish between the two reaction time distributions depicted in red and blue curves in
Fig. \ref{fig:OneSubject_Race1to4}B or Fig. \ref{fig:OneSubject_Race1to4}C but not in 
Fig. \ref{fig:OneSubject_Race1to4}A or Fig. \ref{fig:OneSubject_Race1to4}D.

\section*{Discussion}

\subsection*{The main finding}

Our non-spurious prediction,  ${\rm min}({RT}_{C\!M\!O}, {RT}_C, {RT}_M, {RT}_O) \peq  
{\rm min}({RT}_{C\!M}, {RT}_{C\!O}, {RT}_{M\!O})$,
agrees quantitatively with behavioral data in all six observers such that the distribution of ${RT}_{C\!M\!O}$ can be 
quantitatively predicted from those of the other types of reaction times of the same observer without any free parameters.
This prediction is derived using the following
four essential ingredients: (1) the V1 saliency hypothesis that the highest V1 neural response to a location
relative to the highest V1 responses to other locations signals this location's saliency, (2)
the feature-tuned neural interaction, in particular iso-feature suppression, that depends on the preferred features of
the interacting neurons (e.g., whether the neurons have similar preferred features)
and causes higher responses to feature singletons, (3) the data-inspired assumption that V1
does not have CMO neurons tuned simultaneously to color, motion direction, and orientation,
and (4) the  monotonic link (within the definition of saliency) between a higher saliency of a location and a shorter saliency-dictated reaction time
to find a target at this location.
Hence, our finding supports the direct functional link
between saliency of a visual location and the maximum (rather than, e.g., a summation) of the neural responses
to this location, as prescribed by the V1 saliency hypothesis.
Additionally, it means that saliency computation (at least for our singleton scenes) 
essentially employs only the following neural mechanisms: feature-tuned 
interaction between neighboring neurons (in particular iso-feature suppression) and a lack of CMO neurons, both 
available in V1, and mechanisms which are absent in V1 are not needed.

\subsection*{The supporting findings}

In addition, the following qualitative findings are obtained.
First, two spurious predictions,  ${RT}_{C\!O} \peq {\rm min}({RT}_O, {RT}_O)$ and  ${RT}_{M\!O} \peq {\rm min}({RT}_M, {RT}_O)$, 
about which we have good confidence to be incorrect based on the V1 saliency hypothesis and 
the known presence of the CO and MO cells in V1, are falsified  by our reaction time data.
Second, using the V1 saliency hypothesis and our knowledge about the V1 neural substrates, we predicted relationships 
between the three predictions just mentioned, one non-spuroius and two spurious, and the other five spurious predictions 
listed (in terms of race equalities) in Table 1.
These relationships include the relative degrees of spuriousness between predictions and 
the dependence of some predictions on the non-spuriousness of some other predictions and certain properties of behavioral reaction times.
The outcomes of testing the other five predictions using behavioral reaction times are consistent with the predicted relationships,   
lending further support to the idea that the saliency-dictated behavioral reaction times are indeed
directly linked with the V1 neural responses as prescribed by the V1 saliency hypothesis.

\subsection*{Implications for the V1 saliency hypothesis}

Previously, the V1 saliency hypothesis provided only qualitative predictions. 
An example is the prediction that an ocular singleton should be salient\cite{Zhaoping2008OcularSingleton}, 
another is the prediction that a very salient border between two textures of oblique bars
can be made largely non-salient by a superposed checkerboard pattern of horizontal and vertical bars
(in a way unexpected from traditional saliency models)\cite{ZhaopingMay2007}.
The first one qualitatively predicts that the reaction time to find a visual search target should be shorter when this target is 
also an ocular singleton, but it cannot quantitatively predict how much shorter this reaction time should be.
Similarly, the second one predicts that the reaction time to locate the texture border should be substantially prolonged, 
but not how much prolonged, 
by the presence of the superposing texture.
Confirmations of these qualitative predictions not only support the V1 saliency
hypothesis linking V1 neural responses to behavioral saliency, but also support the idea that (V1) 
neural mechanisms employed in the derivations of the predictions, in particular the iso-feature suppression, 
are used for the saliency computation. However, they cannot conclude whether additional mechanisms 
not yet considered, particularly the more complex mechanisms available only in higher brain centers, 
might also contribute to the saliency computation.
In contrast, if a prediction specifies not only that 
one reaction time should be qualitatively shorter, but also be quantitatively shorter by, say, 20\%,  
than another reaction time, and if data reveal instead that the first reaction time
is only 10\% shorter, then additional mechanisms for saliency computation must be called for.
Now, the quantitative agreement between our non-spurious prediction and the reaction time data without any free parameters 
enables us to conclude that saliency computation requires essentially no other neural mechanisms 
than the feature-tuned interactions between neurons and a lack of CMO neurons --- both are V1 properties.

We should keep in mind that some other mechanistic ingredients or assumptions were omitted in our closing sentence in the last
paragraph. Let us articulate and remind ourselves of these other ingredients which have been explicit or implicit in this paper.
One is the assumption that the response fluctuations in different types of neurons to a single input item are independent of each other.
Hence, for instance fluctuations in the responses of the C, O, and CO neurons to the CO singleton are assumed to be independent of each other.
A related assumption is that the fluctuations of the responses to different input items in a scene are sufficiently independent of each other, so that
we can approximately treat the statistical properties of the responses to the background bars as independent of the responses to 
the singleton.
Another simplification is the assumption that the response of a neuron to a singleton is independent of whether this singleton 
is unique in a feature dimension to which this neuron is not tuned.  For example, we assume that there is no statistical 
difference between $r^C_C$, $r^C_{C\!O}$, and  $r^C_{C\!M\!O}$, or between  
$r^{C\!O}_{C}$ and $r^{C\!O}_{C\!M}$, 
or between $r^C_B$ and $r^C_O$. 
This assumption may not be strictly true given the known activity normalization in cortical responses\cite{Heeger1992}, although it may 
be seen as an approximation.
Of course treating the population responses to the background bars as having the same statistical property 
regardless of the type of the singleton in our singleton scenes is another simplification 
which is in fact only an approximation, it enabled us to write equation (\ref{eq:SingletonSaliency}). 
Meanwhile, equation (\ref{eq:SingletonSaliency}) led to  equation (\ref{eq:SingletonRT}) by an implicit assumption 
that flucutations in the saliency readout to motor responses are negligible (this might be more likely valid for
bottom-up than top-down responses).
Furthermore, we are assuming that perceptual learning by the observers to do visual search is negligible 
over the course of the data taking, so that the monotonic function relating V1 responses to reaction times is fixed.
The above simplifications or idealizations were made to keep our question focused on the
most essential mechanisms.
That our prediction agrees quantitatively with data suggests that the above simplifications or idealizations are
sufficiently good approximations within the resolution that can be discerned by our data.
 
\subsection*{Implications for the role of extrastriate cortices}

An important question is whether extrastriate cortices, i.e., cortical areas beyond V1, might also contribute
to compute saliency.
This question is important, since, if these cortical areas could be excluded from determining saliency,
future investigation of the extrastriate cortices could focus on their role in other functions.
From the discussions in the previous sub-section,  extrastriate cortices could contribute to 
computing saliency if they possess the mechanistic ingredients of feature-tuned interaction (in particular iso-feature suppression) 
between neighboring neurons and a lack of CMO tuned cells.
If so,  we could simply extend the hypothesized link between the highest neural response to a location and
the saliency of this location to extrastriate cortices, which also projects to superior colliculus 
and so can influence eye movements.

It has been known since 1980s\cite{AllmanEtAl1985} that extrastriate cortices also have the feature-tuned
surround interactions,  in particular the iso-feature suppression. 
For example, V4 neurons exhibit iso-color, iso-orientation, and iso-spatial-frequency suppression\cite{DesimoneSchein1987,ScheinDesimone1990},
V2 neurons also exhibit iso-orientation suppression\cite{VanEssenEtAl1989}, and MT neurons 
exhibit iso-motion-direction suppression\cite{AllmanEtAl1985}.

However, extrastriate cortices contain CMO neurons (private communication from Stewart Shipp, 2011).
For example, Burkhalter and van Essen\cite{BurkhalterVanEssen1986} observed that,
in V2 and VP, many cells were feature selective in multiple feature dimensions, including orientation,
color, and motion direction, and that the incidence of selectivity in multiple dimensions was
approximately that which would be expected if the probabilities of occurance of different selectivities
in any given cell were independent of one another. These observations imply that triple-feature  tuned
CMO cells are present even though they are fewer than double- or single-feature tuned cells. In fact, since
they observed that most neurons are tuned to orientation and most neurons are tuned to color, 
the probability that a cell can be a CMO cell must be no less than 25\% 
of the probability of this cell being  tuned to direction of motion (M). 
Similar conclusions in V2 are reached by other investigations\cite{GegenfurtnerEtAl1996, ShippEtAl2009}, 
although the numerical probability of a neuron being a CMO neuron depends  on the criteria to classify whether a neuron is 
tuned to a feature dimension.
In addition, unlike the case in V1 where the presence of CM neurons is controversial, V2 is known 
to have CM neurons, in addition to having CO and MO neurons\cite{GegenfurtnerEtAl1996, TamuraEtAl1996, ShippEtAl2009}.
Some of these CM, CO, and MO neurons (which are defined experimentally as being tuned to the two specified feature 
dimensions simultaneously without restrictions on the neuron's selectivity in the other feature dimensions) in V2 can 
well be CMO neurons, especially when the chance for a neuron to be tuned to another feature dimension is independent 
of the other feature dimensions to which this neuron is already tuned.
Selectivity to conjunctions of more than two types of features in extrastriate cortices
is consistent with general observations that neurons in cortical areas
beyond V1 tend to have more complex and specialized visual receptive fields. 

According to our analysis in the Methods section, if a cortex containing the saliency map 
had CMO neurons, then, statistically,  ${RT}_{C\!M\!O}$ would be likely smaller than
predicted by our non-spurious race equality ${\rm min}({RT}_{C\!M\!O}, {RT}_C, {RT}_M, {RT}_O) \peq  {\rm min}({RT}_{C\!M}, {RT}_{C\!O}, {RT}_{M\!O})$ 
derived from the V1 saliency hypothesis and V1 mechanisms.
That ${RT}_{C\!M\!O}$ would be shorter than predicted is a generalization of the case that 
the presense of CO neurons makes ${RT}_{C\!O}$ shorter than predicted by the race equality  ${RT}_{C\!O}\peq {\rm min}({RT}_C, {RT}_O)$.
More specifically, according to equation (\ref{eq:RTlist=ResponseList}), 
adding the CMO neurons would modify equation (\ref{eq:minRTCMOCMO}) (in the Methods section) 
such that ${\rm min}({RT}_{C\!M\!O}, {RT}_C, {RT}_M, {RT}_O) = f[{\rm max}(Y)]$, where
 $Y$ is a list of responses from the single-feature tuned and double-feature tuned neurons
as specified   in equation (\ref{eq:minRTCMOCMO}), 
 would add four extra items $r^{C\!M\!O}_{C\!M\!O}$, $r^{C\!M\!O}_C$, $r^{C\!M\!O}_M$, and $r^{C\!M\!O}_O$ into the list $Y$.
Similarly, equation ${\rm min}({RT}_{C\!M}, {RT}_{C\!O}, {RT}_{M\!O})=f[{\rm max}(Y)]$ would add extra 
three items $r^{C\!M\!O}_{C\!M}$, $r^{C\!M\!O}_{C\!O}$, and $r^{C\!M\!O}_{M\!O}$ into the same list $Y$.
Consequently, the equality ${\rm min}({RT}_{C\!M\!O}, {RT}_C, {RT}_M, {RT}_O) \peq  {\rm min}({RT}_{C\!M}, {RT}_{C\!O}, {RT}_{M\!O})$, 
which holds when the CMO neurons are absent, would be broken by the presence of CMO neurons unless
either the CMO responses satisfy
\be
{\rm max}\left (r^{C\!M\!O}_{C\!M\!O}, r^{C\!M\!O}_C, r^{C\!M\!O}_M, r^{C\!M\!O}_O \right ) \peq  {\rm max}\left (r^{C\!M\!O}_{C\!M}, r^{C\!M\!O}_{C\!O}, r^{C\!M\!O}_{M\!O} \right ),
 \label{eq:CMOresponseCondition}
\ee
or if the two quantities in both sides of the above equation are negligible compared
to ${\rm max}(Y)$, the maximum response of the list of single- and double-feature tuned neurons.
Since iso-feature suppression would typically make $r^{C\!M\!O}_{C\!M\!O}$ larger than all the other responses
$r^{C\!M\!O}_\alpha$ for $\alpha \ne {C\!M\!O}$, the above equation is likely replaced
by
$\left \la {\rm max}\left (r^{C\!M\!O}_{C\!M\!O}, r^{C\!M\!O}_C, r^{C\!M\!O}_M, r^{C\!M\!O}_O \right ) \right \ra  > \left \la {\rm max}\left (r^{C\!M\!O}_{C\!M}, r^{C\!M\!O}_{C\!O}, r^{C\!M\!O}_{M\!O}\right ) \right \ra$,
 and consequently ${RT}_{C\!M\!O}$ would be smaller than predicted by the race equality unless the CMO responses are immaterial.

Assuming that the responses from the CMO cells in the extrastriate cortices are not negiligible, and assuming that
their responses  under the ubiquitous iso-feature suppression (or more general contextual influences) 
do not satisfy equation (\ref{eq:CMOresponseCondition}) above, then the confirmation of our non-spurious race equality 
${\rm min}({RT}_{C\!M\!O}, {RT}_C, {RT}_M, {RT}_O) \peq  {\rm min}({RT}_{C\!M}, {RT}_{C\!O}, {RT}_{M\!O})$ by our behavioral reaction time data suggests that,
at least for the type of visual inputs that we used in our search task, 
extrastriate cortices contribute little  to the guidance of exogenous attention (excluding the contribution 
to maintaining the state of alertness of observers).
This suggestion is consistent with our previous finding that an eye-of-origin singleton is very salient despite
a paucity of eye-of-origin signals in every cortical area beyond V1.

Meanwhile, as our knowledge about the extrastriate cortices are still sketchy, we cannot rule out the possibility that
the responses of the CMO cells in the extrastriate cortices satisfy equation (\ref{eq:CMOresponseCondition}) above
or are negiligible compared to the responses from cells tuned conjunctively to fewer feature dimensions.
For example, one way to make equation (\ref{eq:CMOresponseCondition}) hold is to have the responses from the
CMO cells invariant to any changes in the contextual inputs outside the classifical receptive fields of these cells, in particular,
to exclude the ubiquitous iso-feature suppression from the CMO cells in the extrastriatex cortices.
The current study hopefully can motivate experimental investigations of the response properties of these cells 
in the extrastriate cortex.

\subsection*{Further discussions assuming no role in saliency by the extrastriate cortices}

Although the current study cannot firmly establish the possibility that extrastriate cortices play no role in the saliency function, 
the implication of such a possibility is so non-trivial that we discuss it  here at the end of this paper.

Traditionally, it has been thought that the control of the direction of attention, including exogenous attention, rests on a network of 
neural circuits comprising frontal and parietal areas, including the frontal eye field 
and intraparietal areas\cite{CorbettaShulman2002,GottliebEtAl1998,IttiKoch2001}.
The role of subcortical areas such as the superior colliculus has also been
suggested\cite{KustovRobinson1996}, although it is likely to merely implement attentional control commands. 
A quantitative exclusion of extra-striate contributions to exogenous control 
should invite a fundamental revision of this network for attentional control.

If extrastriate cortical areas downstream from V1 along the visual pathway 
can be liberated from a role in exogenous attention, they can then focus on 
post-selectional decoding and/or endogenous selection influenced by
top-down goals and expectations.  Furthermore, in light of exogenous attentional control by V1, 
and since attentional selection admits only a tiny fraction of
sensory information to be processed in detail,
visual information processed in the extrastriate cortices is likely to have a much smaller
amount than that fed to V1 from the retina. This consideration should shape our investigations 
and shed light on some past observations. 

Indeed, if we compare  V1 with extrastriate cortical areas,  the neural activities in the former are 
more associated with sensory inputs than perception (i.e., outcomes of visual decoding) and less influenced by top-down attention, 
whereas those in the latter are more associated with perception rather than sensory inputs and 
more influenced by top-down attention\cite{CrickKoch1995}.
For example, lesions in V4 impair visual selection of only non-salient objects\cite{SchillerLee1991} disfavored 
by exogenous selection, demonstrating an involvement of V4 in endogenous selection. 
Equally, neural responses in V4 but not V1 to binocularly  rivalrous inputs are dominated by perceived input rather than the
retinal images\cite{LeopoldLogothetis1996}, contrasting the involvement of V4 and V1 in perceptual decoding.
 
Identifying V1's role in exogenous selection thus helps to crystallize the research questions 
and pave the way for investigating extrastriate cortical areas.

\section*{Methods}

\subsection*{Behavioral data to test various race equalities}

We test predictions of various race equalities using behavioral data previously collected by Koene and Zhaoping\cite{KoeneZhaoping2007}. 
They used dense stimuli, each containing 660 bars, and collected about 300 samples of 
reaction times for each singleton category 
$\alpha = C$, $M$, $O$, $C\!M$, $C\!O$, $M\!O$, or $C\!M\!O$ from each observer.
The observer's task was to find a target bar having a unique feature, regardless of the feature(s)  which distinguished it, and to report
as quickly as possible whether the target was in the left or right half of the display. 
Search trials of different types of singletons were randomly interleaved.
Each stimulus bar was about $1\times 0.2^o$ in visual angle, took one of the two possible
iso-luminant colors (green and purple), tilted from vertical in either clockwise or anticlockwise direction by a
constant amount, and moved left or right at a constant speed. All background bars
were identical to each other in color, orientation,  and motion direction; so
the singleton popped out  by virtue of its unique color, tilt, or motion direction, or any
combination of these features (as schematized in Fig. \ref{fig:SevenSingletons}).  The singleton had an eccentricity $12.8^o$
from the center of the display, which was the
initial fixation point in the beginning of each search trial.
Trials with incorrect button presses or with reaction times shorter than 0.2 seconds or longer than three
standard deviations above the average reaction time (for the particular observer and singleton type)  were
excluded from data analysis. 
Six observers (three of them male) have completed the experiment with  reaction time data on all the seven singleton types. Two 
additional observers (one of them male) however lacked data on ${RT}_{C\!M\!O}$ (since they completed only an earlier version 
of the experiment), hence their data will only be used to test the race equalities not
involving ${RT}_{C\!M\!O}$ (these equalities were the focus of Koene and Zhaoping's study).  
More details about the experiment can be found in the original paper\cite{KoeneZhaoping2007}, 
which did not publish or use the ${RT}_{C\!M\!O}$ data.

The behavioral experiment was designed such that there is a symmetry between the two distinct feature values in any feature dimension,
C, O, or M. For example, the two color features, green and purple, are equally luminant, so that it is reasonable to assume that
the two C singleton stimuli, one is a green bar in a background of purple bars and the other is a purple bar in a background
of green bars, evoke the same population response levels at least in a statistical sense.
More explicitly, we assume that the response level to the color singleton is drawn from the same distribution regardless of whether
the singleton is green or purple, even though the most responsive neurons to the two singletons differ in their color preference.
Furthermore, it is also reasonable to assume that the population responses to the background bars are statistically the same so 
that the two stimuli share the same invariant background response distribution, even though the two backgrounds activate 
different neural populations.  Then, we can treat the two color singleton stimuli the same in terms of 
saliency, which is feature-blind once the response levels are given.  
Therefore, given an observer, our data analysis pools all the ${RT}_C$ data samples into a single pool regardless of 
whether the singleton is green or purple. 
Analogously, it is reasonable to assume that all singleton scenes share the same invariant background response distribution 
regardless of the singleton type, a singleton scene is distinguished by whether it is a C, M, O, CM, CO, MO, or CMO 
singleton scene regardless of the feature values of the singleton and background bars. 
Hence, for each observer and given a 
$\alpha = C$, $O$, $M$, $C\!M$, $C\!O$, $M\!O$, or $C\!M\!O$, 
we pool all this observer's  ${RT}_\alpha$ data samples into a single pool for data analysis regardless of the feature values 
of any input bars.

\subsection*{Proof of the non-spurious race equality in equation (\ref{eq:CMOrace})}

To prove this equality between the two races,
${\rm min}({RT}_{C\!M\!O}, {RT}_C, {RT}_M, {RT}_O)$  
and ${\rm min} ({RT}_{C\!M}, {RT}_{C\!O}, {RT}_{M\!O})$,
 we use equation (\ref{eq:RTlist=ResponseList}) to write
both races in terms of $f[{\rm max}(...)]$. 
The equality holds when both races can be expressed by the 
same list of neural responses as the arguments in $f[{\rm max}( ....)]$.
To start, we express, like we did for ${RT}_C$ and ${RT}_O$  in equation (\ref{eq:RTcAndRTo}), 
each racer's reaction time by 
\begin{equation}
\mbox{reaction time for a singleton} = f[{\rm max}(\mbox{list of non-trivial neuron responses to the singleton})].
\end{equation}
First, we generalize to six types of V1 neurons, three tuned to single features
C, M, and O and three to the three combinations CM, CO, and MO, and none tuned to CMO.
Second, we generalize to seven types $\alpha$ of singleton bars shown in Fig. \ref{fig:SevenSingletons}, 
three single-feature singletons, $\alpha = C$, $M$, and $O$,
three double-feature singletons, $\alpha = C\!M$, $C\!O$, and $M\!O$, and one triple-feature singleton $\alpha = C\!M\!O$.
Each type of singleton evokes responses from all six types of neurons (the preferred feature of each type of neuron matches
the relevant feature of the bar). The response of each neuron type $X = C$, $M$, $O$, $C\!M$, $C\!O$, or $M\!O$ to a singleton type 
$\alpha = C$, $M$, $O$, $C\!M$, $C\!O$, $M\!O$, or $C\!M\!O$, or even to a background bar $\alpha = B$, is denoted as 
$r^X_\alpha$.
For example, by equation (\ref{eq:SingletonRT}) and analogous to equation (\ref{eq:RTc}) we have 
\begin{eqnarray}
{RT}_C  &=& f\left [{\rm max}\left (r^C_C, r^M_C, r^O_C, r^{C\!M}_C, r^{C\!O}_C, r^{M\!O}_C \right ) \right ] \label{eq:RTcFirstline}\\
      &=& f\left [{\rm max}\left (r^C_C, r^M_B, r^O_B, r^{C\!M}_C, r^{C\!O}_C, r^{M\!O}_B \right ) \right ]\label{eq:RTcSecondline}. 
\end{eqnarray}
For the second line above, we used equalities $r^M_C=r^M_B$, $r^O_C=r^O_B$, and $r^{M\!O}_C=r^{M\!O}_B$  which,  analogous to 
equations (\ref{eq:CBCO}--\ref{eq:OBOC}), arise because (due to iso-feature suppression)
a neuron's responses to a singleton bar and a background bar are the same
unless the singleton is unique in at least one the feature dimensions to which this neuron is tuned.
Then, analogous to equation (\ref{eq:RTcAndRTo}), we can ignore all the trivial response levels $r^X_B$ to get
\begin{equation}
{RT}_C = f\left [{\rm max}\left (r^C_C, r^{C\!M}_C, r^{C\!O}_C \right ) \right ].
\end{equation}	
Hence, ${RT}_C$ is determined by only the non-trivial neural responses 
$r^C_C$, $r^{C\!M}_C$, and $r^{C\!O}_C$.
In Fig. \ref{fig:SevenSingletons}, these three non-trivial responses
are listed under the schematic for the C singleton.
Analogously,  we have
\begin{eqnarray}
        {RT}_M  &=& f\left [\mm \left (r^C_M, r^M_M, r^O_M, r^{C\!M}_M, r^{C\!O}_M, r^{M\!O}_M\right )\right ] = f\left [\mm\left (r^M_M, r^{C\!M}_M, r^{M\!O}_M \right ) \right ], \nonumber  \\
        {RT}_O  &=& f\left [\mm \left (r^C_O, r^M_O, r^O_O, r^{C\!M}_O, r^{C\!O}_O, r^{M\!O}_O\right )\right ] = f\left [\mm\left (r^O_O, r^{C\!O}_O, r^{M\!O}_O \right ) \right ]. \nonumber 
\end{eqnarray}
For the reaction time ${RT}_{C\!M}$ for the double-feature singleton CM, we have
\begin{eqnarray}
        {RT}_{C\!M}  &=& f\left [\mm \left (r^C_{C\!M}, r^M_{C\!M}, r^O_{C\!M}, r^{C\!M}_{C\!M}, r^{C\!O}_{C\!M}, r^{M\!O}_{C\!M} \right ) \right ] \nonumber  \\
                 &=& f\left [\mm \left (r^C_{C},  r^M_{M},  r^O_{B},  r^{C\!M}_{C\!M}, r^{C\!O}_{C},  r^{M\!O}_{M} \right ) \right ] \nonumber  \\
                &=& f\left [\mm\left (r^C_C, r^M_M, r^{C\!M}_{C\!M}, r^{C\!O}_C,  r^{M\!O}_M \right ) \right ]. \nonumber 
\end{eqnarray}
The second line above used equality $r^C_{C\!M}= r^C_C$, $r^M_{C\!M} = r^M_M$, $r^O_{C\!M} = r^O_B$, $r^{C\!O}_{C\!M} = r^{C\!O}_C$, and $r^{M\!O}_{C\!M}=r^{M\!O}_M$ which arise
by the same or analogous reason behind the equalities $r^M_C=r^M_B$, $r^O_C=r^O_B$, and $r^{M\!O}_C=r^{M\!O}_B$ used to derive equation 
(\ref{eq:RTcSecondline}) from equation (\ref{eq:RTcFirstline}), namely, a neuron equates a unique feature with a background 
feature unless the neuron is tuned in this feature dimension.
Analogously,
\begin{eqnarray}
        {RT}_{C\!O}  &=& f\left [\mm \left (r^C_{C\!O}, r^M_{C\!O}, r^O_{C\!O}, r^{C\!M}_{C\!O}, r^{C\!O}_{C\!O}, r^{M\!O}_{C\!O} \right ) \right ]  \nonumber  \\
                &=& f\left [\mm\left (r^C_C, r^O_O, r^{C\!M}_C, r^{C\!O}_{C\!O},  r^{M\!O}_O \right ) \right ]  \quad\mbox{and}\nonumber  \\
        {RT}_{M\!O}  &=& f\left [\mm \left (r^C_{M\!O}, r^M_{M\!O}, r^O_{M\!O}, r^{C\!M}_{M\!O}, r^{C\!O}_{M\!O}, r^{M\!O}_{M\!O} \right ] \right )  \nonumber  \\
                &=& f\left [\mm\left (r^M_M, r^O_O, r^{C\!M}_M, r^{C\!O}_{O},  r^{M\!O}_{M\!O} \right ) \right ]. \nonumber  
\end{eqnarray}
Similarly, again treating a unique feature as a background feature for any neuron not tuned to the corresponding feature dimension, we have
\begin{eqnarray}
        {RT}_{C\!M\!O}  &=& f\left [\mm \left (r^C_{C\!M\!O}, r^M_{C\!M\!O}, r^O_{C\!M\!O}, r^{C\!M}_{C\!M\!O}, r^{C\!O}_{C\!M\!O}, r^{M\!O}_{C\!M\!O} \right ) \right ] \nonumber  \\
                &=& f\left [\mm\left (r^C_C, r^M_M, r^O_O, r^{C\!M}_{C\!M}, r^{C\!O}_{C\!O}, r^{M\!O}_{M\!O} \right ) \right ]. \nonumber 
\end{eqnarray}
In Fig. \ref{fig:SevenSingletons}, the non-trivial responses  to determine each singleton's reaction time 
are listed under the corresponding
schematic.

Using six types of V1 neurons (C, M, O, CM, CO, MO) instead of three types of V1 neurons (C, O, CO), one can
revise the derivations in equations (\ref{eq:RTc}--\ref{eq:RaceRTcAndRTo}) to verify that
the race equality 
${RT}_{C\!O}
\peq {\rm min}({RT}_C, {RT}_O)$
 still does not hold in general. 

Now, we apply equation (\ref{eq:RTlist=ResponseList}) to arrive at the race at the left-hand side of equation (\ref{eq:CMOrace}) as
\begin{eqnarray}
{\rm min} 
 &&\!\!\! \!\!\!\!\!\!\!\!\!\!\!\!\! 
({RT}_{C\!M\!O}, {RT}_C, 
{RT}_M, {RT}_O ) \nonumber \\
	&=&f\left [\mm \left(r^C_C, r^C_C, r^O_O, r^O_O, r^M_M, r^M_M,  
		r^{C\!M}_C, r^{C\!M}_M, r^{C\!M}_{C\!M}, r^{C\!O}_C, r^{C\!O}_O, 
	r^{C\!O}_{C\!O}, r^{M\!O}_M, r^{M\!O}_O, r^{M\!O}_{M\!O} \right)\right].
\label{eq:minRTCMOCMO}
\end{eqnarray}
One can easily verify that the list of the arguments in the $f[{\rm max}( ...)]$ 
above is the collection of all the non-trivial neural responses listed under the corresponding
singleton stimuli in Fig. (\ref{fig:SevenSingletons}).  
Similarly, the race ${\rm min}({RT}_{C\!M}, {RT}_{C\!O}, {RT}_{M\!O})$ at the right-hand side of  equation (\ref{eq:CMOrace}) gives the 
same outcome as above, thus proving the race equality.
Again, this equality holds regardless of the details in the form of the saliency read-out function $f(.)$
as long as this function is monotonically decreasing. 

Note that in the expression $f[{\rm max}( ...)]$ in equation (\ref{eq:minRTCMOCMO}), each of $r^C_C$, $r^O_O$ and $r^M_M$ occurs twice.
The expression should not be simplified by deleting the repetitions
 because the neural responses are stochastic. For example, the two occurances of $r^C_C$ should be understood 
as two independent and random samples of $r^C_C$ from its probability distribution.
If $r^C_C$ follows a Poisson distribution with an average of $20$ spikes/second,  the two occurances of $r^C_C$ jointly contribute to the race
by the maximum value of the two random samples from this distribution, and this maximum value is on average larger than $20$ spikes/second.

\subsection*{Methods to test a race equality}

\begin{figure}[hhhhhhthh!!]
\begin{center}
\includegraphics[width=145mm]{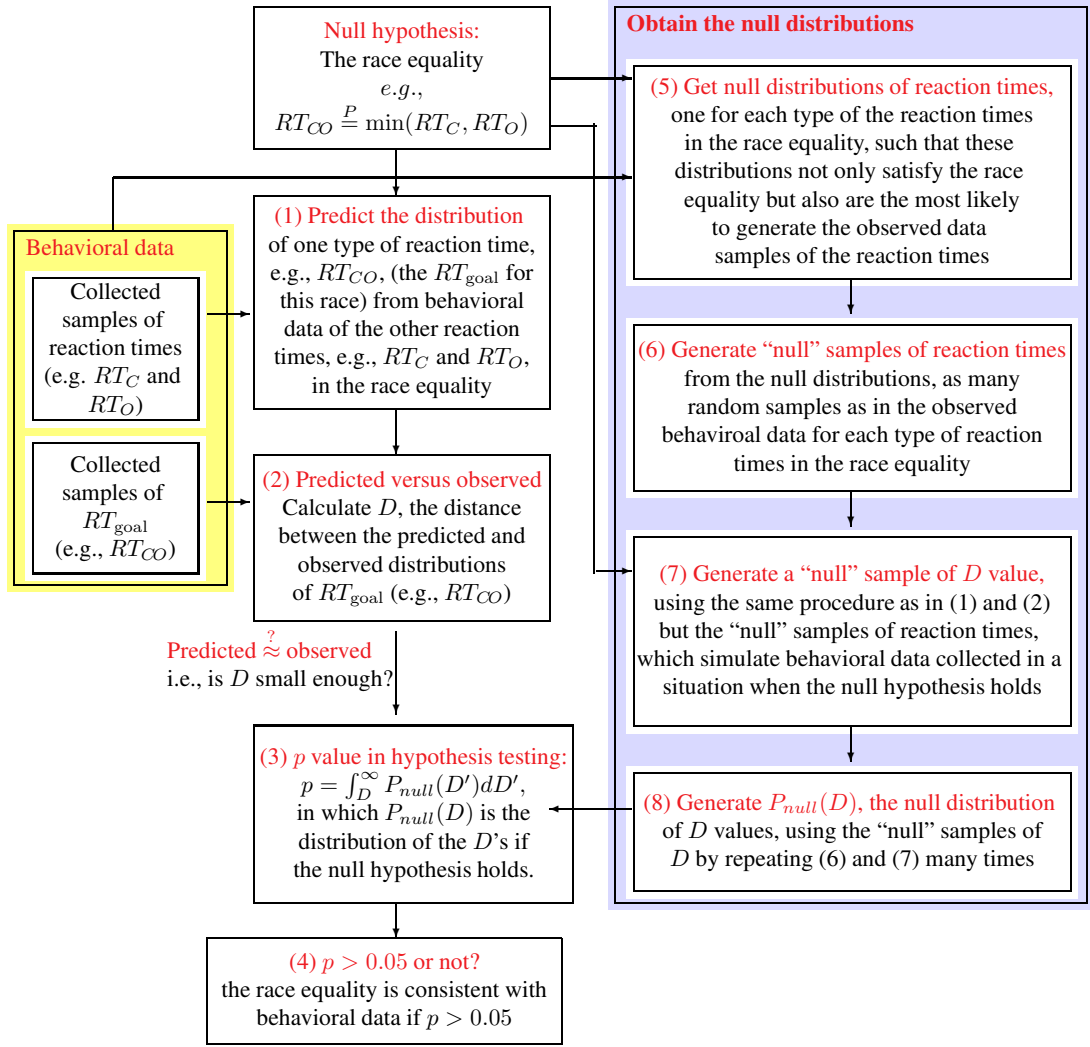}
\end{center}
\caption{ \label{fig:TestingMethod}
Diagram to outline the methods used to test each  race equality, using $RT_{C\!O}\peq {\rm min}(RT_C, RT_O)$ as an example.
This diagram also applies to other race equalities. The details of various components, in boxes (1)-(8) of this figure, 
are described in the text.
}
\end{figure}

Fig. \ref{fig:TestingMethod} outlines our methods to test each race equality against behavioral data, using the equality
$RT_{C\!O}\peq {\rm min}(RT_C, RT_O)$ as an example.
Briefly, given a race equality, the distribution of $RT_{\rm goal}$, the designated type of reaction times in a given equality 
(e.g., $RT_{C\!O}$ is the $RT_{\rm goal}$ in race equality $RT_{C\!O}\peq {\rm min}(RT_C, RT_O)$), 
is predicted from the behaviorally observed distributions of the other reaction times in this equality.
The predicted distribution is compared with the behaviorally observed distribution of $RT_{goal}$, and a distance
$D$ between these two distributions is calculated. Ideally, a zero $D$ means that the race equality agrees with data.
However, this distance $D$ is typically non-zero even when a race equality does hold, since finite numbers of reaction time data samples can only
approximately represent the underlying distributions of various reaction times.
A statistical test is devised to give a $p$ value in testing the null hypothesis of the race equality, such that $p$ is
the probability that the distance $D$ between the predicted and observed $RT_{\rm goal}$  should be at least as big as observed if the
race equality holds. A $p>0.05$ is chosen to suggest that the race equality is consistent with behavioral data.
Our testing methods involve several components which are represented by various boxes in Fig. \ref{fig:TestingMethod}.
The rest of the Methods section describes the details in each of these components for interested readers.

\subsubsection*{Methods to predict a distribution of reaction times from a race equality}

Here we describe the method component in box (1) of Fig. \ref{fig:TestingMethod}.
A race equality enables us to predict the distribution of  one type of reaction times in the race equality from
those of the other reaction times in the race equality.
The predicted reaction time,  denoted by
${RT}_{\rm goal}$ , is designated as 
${RT}_{C\!M\!O}$ in all race equalities except $RE_i$ for $i=2$--$4$ for which
${RT}_{\rm goal}= RT1={RT}_{C\!O}$, ${RT}_{M\!O}$,  or ${RT}_{C\!M}$, 
respectively, as listed in Table 1.

In $RE_i$ for $i=2$--$5$, the distribution of ${RT}_{\rm goal} = RT1$ can be predicted directly
as the distribution of the race outcome $RT2$.
Given any  race, e.g.,  ${\min }({RT}_C, {RT}_O)$ or a three-racer race $RT2\equiv {\rm min}({RT}_{C}, {RT}_{M}, {RT}_{O})$, the samples or the distribution of $RT2$, the race winner, can be obtained by
the following method:
\begin{eqnarray}
\mbox{all winner RT samples of a race}&=&\mbox{the minimums of all possible combinations of  } \nonumber \\
		&&\mbox{the reaction time samples from the racers.} \label{eq:RaceModel}
\end{eqnarray}
This predicted ${RT}_{\rm goal}$ distribution can then be compared with the distribution of the collected behavioral data samples 
of ${RT}_{\rm goal}$. 

The predictions of $RT_{\rm goal}$ in $RE_1$ and $RE_i$ for $i=6$--$8$ use a different and more complex method.
In these races, the $RT_{\rm goal}$ is always 
${RT}_{C\!M\!O}$, and  $RT1$ can be written as $RT1 \equiv {\rm min}({RT}_{C\!M\!O}, {RT}_{\rm part})$, where 
$ {RT}_{\rm part}$ is the winner reaction time from another race involving only the 
racers in the $RT1$ race other than the racer ${RT}_{C\!M\!O}$.
Explicitly, for $RE_1$, $RE_6$, $RE_7$, or $RE_8$, 
respectively,
${RT}_{\rm part} = {\min}({RT}_C, {RT}_M, {RT}_O)$, 
${RT}_{\rm part} = {\min}({RT}_M, {RT}_{C\!O})$, 
${RT}_{\rm part} = {\min}({RT}_C, {RT}_{M\!O})$, or
${RT}_{\rm part} = {\min}({RT}_O, {RT}_{C\!M})$.
The samples of $RT_{\rm part}$ can be obtained by using equation (\ref{eq:RaceModel}). 
For each observer and each race equality,  the samples of 
${RT}_{C\!M\!O}$, ${RT}_{\rm part}$, and $RT2$ are
discretized into $N$ time bins bounded by time values $t_0< t_1< ... < t_{N}$. 
These time bins are common for ${RT}_{C\!M\!O}$, ${RT}_{\rm part}$, and $RT2$, 
and different $t_i$'s are (in most data analysis) roughly evenly spaced except for very small and 
large $t_i$'s.  Since each observer had about $300$ samples of reaction times for each singleton type, 
the value $N\approx 10$ is chosen in order to give a sufficiently large number of behavioral data samples in each 
time bin while still providing a sufficiently large $N$ to build a reaction time distribution ($N=8-12$ have been tried to test the robustness of our conclusion to these details).
For any particular $RT_\alpha$ (with $\alpha$ denoting a singleton type or a race winner),  
if $n_i$ is the number of the $RT_\alpha$ samples 
in the $i^{th}$ time bin, the distribution of
$RT_\alpha$ across the time bins  is described by an $N$-dimensional vector whose $i^{th}$ component is $n_i/(\sum_j n_j)$.

Let  $N$-dimensional vectors ${\bf P} \equiv (P_1, P_2, ..., P_N)$ and ${\bf Q} \equiv (Q_1, Q_2, ..., Q_N)$ denote the distributions of
 $RT1$ and $RT2$, respectively, in these time bins, and 
 let ${\bf p}$ and ${\bf q}$ denote the distributions of ${RT}_{C\!M\!O}$ and  ${RT}_{\rm part}$, respectively.
$RT1 \equiv {\rm min}({RT}_{C\!M\!O}, {RT}_{\rm part})$ 
means
\begin{equation}
	P_i = p_i (1- \sum_{j\le i} q_j) + q_i (1-\sum_{j\le i} p_j) +  p_i q_i, \quad \mbox{for all $i$}. \label{eq:Pequation}
\end{equation}
Then  $RT1\peq RT2$  means $P_i = Q_i$, i.e.,
\begin{equation}
	 p_i (1- \sum_{j\le i} q_j) + q_i (1-\sum_{j\le i} p_j) +  p_i q_i = Q_i, \quad \mbox{for all $i$}. \label{eq:PeqQ}
\end{equation}

From reaction time data on ${RT}_C$, ${RT}_M$, ${RT}_O$, ${RT}_{C\!M}$, ${RT}_{C\!O}$, 
and ${RT}_{M\!O}$, the samples for ${RT}_{\rm part}$ and $RT2$ can be obtained
by equation (\ref{eq:RaceModel}) to construct the distributions ${\bf q}$ and ${\bf Q}$. 
Then, ${\bf p}$ can be predicted as the
solution to the above linear equation of ${\bf p}$, provided that this solution satisfies the probability
constraints $p_i \ge 0$ and $\sum_i p_i=1$.  
If the solution violates $p_i \ge 0$ or $\sum_i p_i=1$ (this can happen
for example when $q_i > Q_i$ for some $i$ due to sampling noise arising from the limited data samples and/or due to 
a lack of actual race equality in reality), then the predicted ${\bf p}$ is chosen as the one that 
minimizes a distance between ${\bf P}$ and ${\bf Q}$ while satisfying the constraints  $p_i \ge 0$ and $\sum_i p_i=1$ 
(through an optimization procedure, e.g., via the fmincon routine in MATLAB).
The following four different distance measures (between ${\bf P}$ and ${\bf Q}$) were separately tried to test the 
robustness of our conclusion in the paper
\begin{equation}
\begin{array}{ll}
(1): & |{\bf P} - {\bf Q}|^2, \quad \mbox{~the squared Hemming distance,}    \\
(2): & \sum_i ( \sqrt{P_i} -  \sqrt{Q_i})^2,\quad   \mbox{~the Hellinger distance,} \\
(3): & \sum_i | P_i -  Q_i|,   \quad \mbox{the 1-norm distance, and} \\
(4) & \sum_i {\rm max}(Q_i, \epsilon) \log { {{\rm max}(Q_i, \epsilon )}\over {{\rm max}(P_i, \epsilon )} }, 
\mbox{~~~ with a given $\epsilon \ll 10^{-100}$.} 
\end{array}
\label{eq:FourDistances}
\end{equation}
The last distance is the Kullback-Leibler divergence if all $P_i$ and $Q_i$ were larger than a very small $\epsilon$.

Unless  interested in repeating the procedure in this method, readers may wish to skip the rest of this paragraph which
describes how $t_i$'s are determined for each race equality $RE_j$ for $j=1,2,...,8$.
Given a subject and a race equality, all the behavioral data samples of the reaction times for all the singleton types
involved in this race equality are put into a single pool.
This pool of samples were divided into $L=100$ time bins bounded by time boundaries denoted 
as $T_i$'s,  ordered as, 
\begin{equation}
T_0 < T_1 < T_2 < ...< T_L,  \label{eq:Tlist}
\end{equation}
whose values are chosen such that all bins contain (as close as possible) an equal number of 
samples of reaction times from this pool.  
For reasons that will be clear in the next method section, 
each $t_i$ is chosen from among these $T_i$'s as follows.
Let $RT({\rm max})$ and $RT({\rm min})$ denote the
largest and smallest reaction times, respectively, from all the behavorial data samples of
${RT}_{\rm goal}$, $RT2$, and (for $RE_1$ and $RE_i$ for $i=6$--$8$) ${RT}_{\rm part}$.
Given $(T_0, T_1, ...,  T_{L})$,  
$t_0$ is the largest $T_j$ smaller than $RT({\rm min})$ and
$t_N$ is the smallest $T_j$ larger than $RT({\rm max})$.
Then, let $RT'({\rm max})$ and $RT'({\rm min})$ denote the
largest and smallest ${RT}_{\rm goal}$ behavioral data samples, respectively.
If $RT'({\rm min }) > RT({\rm min })$ and the largest $T_j$ smaller than  $RT'({\rm min})$ is larger than
$t_0$, then this $T_j$ is assigned to $t_1$.
If $RT'({\rm max}) < RT({\rm max})$ and the smallest  $T_j$ larger than  $RT'({\rm max})$ is smaller than
$t_N$, then this $T_j$ is assigned to $t_{N-1}$.
Depending on whether  $t_1$ and $t_{N-1}$ have just been assigned, there are now 
$N' = N-1$, $N-2$, or $N-3$ of the unassigned $t_i$ values, which will
be assigned in ascending order to $\tau_1< \tau_2 < ...< \tau_{N'}$.
Each $\tau_i$ is the $T_j$ value not yet assigned to 
$\tau_k$ for $k<i$ and is closest to the value $\tau'_i$ which 
is larger than a fraction $F_i$ (with $F_1< F_2< ...< F_{N'}$) of the ${RT}_{\rm goal}$  data samples. 
Our data analysis tried each of the following four ways to choose $F_i$'s 
to see whether this paper's  conclusion is robust against variations in these details.
One is to choose $F_i = i/(N'+1)$.  The other three uses 
\begin{equation}
F_i = \left (erf \left (-x_F + 2x_F {{i-1}\over {N'-1}} \right )+1\right )/2, \label{eq:ChooseXF}
\end{equation}
in which  $erf(.)$ is the error function and $x_F>0$ is a parameter with value
$x_F=1.25$, $1.35$, or $1.45$.

\subsubsection*{The statistical test for the hypothesis that the predicted and the behavoirally observed distributions arise 
from the same underlying entity}
\label{sec:Test}

We cannot use the Kolmogorov-Smirnov test to see whether $RT1$ samples and $RT2$ samples are generated
from the same underlying distribution, because the samples of at least $RT2$ are not independently generated 
due to the underlying race between the racers. Hence, we devised the following  statistical test to test whether
the predicted and observed distributions of ${RT}_{\rm goal}$ arise from the same underlying entity.
This section details the methods in boxes (2)-(8) of Fig. \ref{fig:TestingMethod}.

The method (box (2) of Fig. \ref{fig:TestingMethod}) to measure the distance $D$ between the predicted and observed distributions is as follows.
Given an observer and a race equality, let ${\bf p}$ and ${\bf \tilde p}$ denote the
predicted and observed distributions of ${RT}_{\rm goal}$ (in the time bins used for predicting the distribution of the reaction times), 
respectively. Let $D$ denote the difference between them.  This difference is measured by one of the four distance metrics as listed 
in equation (\ref{eq:FourDistances}), substituting  ${\bf \tilde p}$ and ${\bf p}$ for ${\bf P}$ and ${\bf Q}$, respectively.  
All the metrics have been tried to test the robustness of our conclusion.

This paragraph describes the methods associated with boxes (3), (4), (7), and (8) of  Fig. \ref{fig:TestingMethod}. 
To test whether $\tilde {\bf p}$ and ${\bf p}$ are statistically indifferent from each other, 
we generated $m=500$ other, simulated, distances $D$, 
each from a set of simulated samples of reaction times (there are as many simulated samples as in the real behavioral data for each type of reaction time) 
collected from a simulated behavioral experiment in a hypothetical situation when the race equality holds while the 
distribution of the simulated samples of reaction times resembles that of the real behavioral reaction time samples.
Given the fixed time boundaries $T_0<T_1<T_2< ... < T_L$ obtained from the
real behavioral data, the procedure to obtain a (simulated) $D$ value using the simulated samples of reaction times 
is the same as that to get a real $D$ value when the real reaction time samples are used.
The $p$ value of the statistical test is the fraction of the simulated $D$ values which are larger than the real $D$ value 
(obtained using the real  behavioral data), a $p<1/m = 0.002$ is given when no simulated $D$ is larger than the real $D$.
Our predicted and observed distributions of ${RT}_{\rm goal}$ are said to be significantly different
from each other, i.e., not arising from the same underlying entity, and we declare that the race
equality is not consistent with behavioral data, when $p<0.05$.  

Each set of simulated samples of reaction times (for a given race equality),
serving as data collected in a simulated behavioral experiment in a hypothetical situation when the race equality holds,
 is generated  as follows (box (6) of Fig. \ref{fig:TestingMethod}). 
First, we should have already constructed (detailed in the next paragraph) a set of probability distributions of the corresponding 
set of reaction times involved in a race equality. 
For example, for race equality $RT_{C\!O}\peq {\rm min}(RT_C, RT_O)$, we should have already available a set of three distributions, one each for
$RT_{C\!O}$, $RT_C$, and $RT_O$, respectively.
This set of distributions is such that, first, it actually satisfies the race equality and, second, given the constraint that the race equality is satisfied, 
the distributions are the most likely to be the underlying distributions from which the behaviorally observed samples of reaction times could be generated.
These distributions are called the null distributions of reaction times for the race equality concerned.
From each of these distributions, as many simulated samples of reaction times as the corresponding real behavioral samples of reaction times 
(for a particular singleton type) are generated as random samples.

The null distributions of reaction times are constructed as follows (box (5) of Fig. \ref{fig:TestingMethod}).
Given a subject and a race equality, the real behavioral reaction times
for all singleton types involved in the race equality 
are discretized into $L=100$ time bins using time boudnaries $T_0< T_1< ... < T_L$ 
as described in and around equation (\ref{eq:Tlist}).
For each singleton type $\alpha$,  
let ${\bf n}_\alpha \equiv  [(n_{\alpha})_1, (n_{\alpha})_2, ...,(n_{\alpha})_L]$ be the
histogram of the real behavioral ${RT}_{\alpha}$ samples in these time bins. 
The likelihood, or probability, that 
an underlying distribution 
$\hat {\bf p}_\alpha  \equiv (\hat p_{\alpha_1}, \hat p_{\alpha_2}, ..., \hat p_{\alpha_L})$
of the reaction times over these time bins is the generator of this histogram ${\bf n}_\alpha$ is
\be
	{\rm likelihood} (\hat {\bf p}_\alpha)  \propto \Pi^L_{i=1} \left (\hat p_{\alpha_i}\right )^{n_{\alpha_i}},
\ee
	whose logarithm is
\be 
\ln \left [{\rm likelihood}(\hat {\bf p}_\alpha) \right ] = \sum_{i=1}^L 
			n_{\alpha_i} \ln \hat p_{\alpha_i} +\mbox{constant}.
\ee
We construct hypothetical probability distributions, $\hat {\bf p}_\alpha$, one for each $\alpha$ 
of the singleton types involved in the race equality,
such that the log-likelihood
\be
	\sum_{\alpha } \ln ({\rm likelihood}(\hat {\bf p}_\alpha)) = \sum_\alpha \sum_{i=1}^L n_{\alpha_i} \ln \hat p_{\alpha_i} +\mbox{constant} \label {eq:SumAlpha}
\ee
is maximized, subject to the constraints that the race equality $RT1 \peq RT2$ (which takes
the form like equations (\ref{eq:Pequation}--\ref{eq:PeqQ}))
is satisfied by these $\hat {\bf p}_\alpha$s and, 
for each $\alpha$, $\sum_{i=1}^L \hat p_{\alpha_i} = 1$ and  $\hat p_{\alpha_i} \ge 0$.  
Again, 
this can be achieved through an optimization procedure (e.g., using fmincon in MATLAB).  We verified that
the resulting $\hat {\bf p}_\alpha$s indeed satisfy the race equality $RT1 \peq RT2$ and sufficiently 
resemble the respective histograms of behavioral data ${RT}_\alpha$. 
Then, for each singleton type $\alpha$, a probability distribution of reaction times over continuous time duration
$(T_0, T_L)$ is constructed from $\hat {\bf p}_\alpha$ such that, 
the probability density within the time window $[T_{i-1}, T_i)$ 
is uniform and equal to $\hat  p_{\alpha_i}/(T_i-T_{i-1})$.

Note that since the time boundaries, the $t_i$'s, for the 
coarser  time bins used in predicting the distribution of $RT_{\rm goal}$  (box (1) of Fig \ref{fig:TestingMethod}) 
are chosen from among the time boundaries, the $T_j$'s, for the finer time 
bins for the null probability distributions which
strictly satisfy the race equality, the race equality remains satisfied when each $\hat p_\alpha$ is viewed
through the coarser time bins used for predicting the ${RT}_{\rm goal}$  distribution.

\section*{Acknowledgement} This work is supported by the Gatsby Charitable Foundation and the 
Tsinghua University 985 fund. We like to thank Peter Dayan, Peter Latham, and Keith May for careful reading of the manuscript 
with very helpful comments for presentations. 


\end{document}